\documentclass[reprint,
 amsmath,amssymb,
 aps,
]{revtex4-2}
\usepackage{graphicx}
\usepackage{dcolumn}
\usepackage{xcolor}
\usepackage{setspace}
\usepackage{cases}
\usepackage{bm}
\usepackage{enumitem}
\usepackage{booktabs}
\usepackage{textcomp}
\usepackage{makecell}
\usepackage{multirow}

\usepackage{xr}
\newcommand{\be}{\begin{equation}}
\newcommand{\ee}{\end{equation}}
\newcommand{\bg}{\begin{equation}}
\newcommand{\eg}{\end{equation}}
\newcommand{\bdm}{\begin{displaymath}}
\newcommand{\edm}{\end{displaymath}}
\newcommand{\bea}{\begin{eqnarray}}
\newcommand{\eea}{\end{eqnarray}}
\newcommand{\beas}{\begin{eqnarray*}}
\newcommand{\eeas}{\end{eqnarray*}}
\newcommand{\ba}{\begin{array}}
\newcommand{\ea}{\end{array}}
\newcommand{\nn}{\nonumber}

\newcommand{\bfg}{\begin{figure}}
\newcommand{\efg}{\end{figure}}

\newcommand{\fr}{\frac}

\newtheorem{lm}{Lemma}
\newtheorem{cl}{Corollary}
\newtheorem{df}{Definition}

\newcommand{\blm}{\begin{lm}}
\newcommand{\elm}{\end{lm}}
\newcommand{\bcl}{\begin{cl}}
\newcommand{\ecl}{\end{cl}}
\newcommand{\bdf}{\begin{df}}
\newcommand{\edf}{\end{df}}
\newcommand{\brk}{\begin{rm}}
\newcommand{\erk}{\end{rm}}

\newcommand{\DD}{{\cal D}}

\newcommand{\Om}{\Omega}

\newcommand{\vE}{{\bf E}}

\newcommand{\vF}{{\bf F}}

\newcommand{\vH}{{\bf H}}

\newcommand{\vJ}{{\bf J}}

\newcommand{\vp}{{\bf p}}

\begin{document}

\title{Harnessing superdirectivity in dielectric spherical multilayer antennas}

\author{Roman Gaponenko,$^{1}$ Alexander Moroz,$^2$ Ilia Rasskazov,$^{3,*}$ Konstantin Ladutenko,$^1$ Alexey Shcherbakov,$^{1,*}$ and Pavel Belov$^1$}
\affiliation{%
 $^1$ School of Physics and Engineering, ITMO University, 197101, Saint-Petersburg, Russia \\ $^2$ Wave-scattering.com \\ $^3$ The Institute of Optics, University of Rochester, Rochester, New York 14627, USA \\
 * Corresponding authors: irasskaz@ur.rochester.edu; alexey.shcherbakov@metalab.ifmo.ru
}%

\begin{abstract}
Small form-factor, narrowband, and highly directive antennas are of critical importance in a variety of applications spanning wireless communications, remote sensing, Raman spectroscopy, and single photon emission enhancement.
Surprisingly, we show that the classical directivity limit can be appreciably surpassed for electrically small multilayer spherical antennas excited by a point electric dipole even if limiting ourselves to purely dielectric materials.
Experimentally feasible designs of superdirective antennas are established by using a stochastic optimization algorithm combined with a rigorous analytic solution.
\end{abstract}
\maketitle

\section{Introduction}
\label{sec:intro}
Physical limitations on antennas are critical when designing efficient devices. The limitations have been the subject of research since the early age of antenna science \cite{bouwkamp_problem_1945,chu_physical_1948,riblet_note_1948}.
There is a trade-off between gain, $G$, which defines the spatial narrowness of radiation patterns, and the quality factor, $Q$, which characterizes the ratio of the stored energy to emitted power~\cite{hansen_fundamental_1981}. $Q$ is known to be inversely proportional to the bandwidth when $Q\gg 1$ \cite{hansen_fundamental_1981}. Gain is related to directivity, $\mathcal D$, through $G=e_r\mathcal D$, where $e_r$ is the radiation efficiency of the antenna, taking into account antenna losses (it does not include polarization or impedance mismatch losses).
In our paper, we consider purely dielectric antennas.
Given that such antennas are lossless ($e_r = 1$), we use only the term {\em directivity} in what follows without any loss of generality.

The Harrington-Chu limit~\cite{chu_physical_1948,Harrington1958} puts an upper bound on the directivity as $\mathcal D_{\rm lim} =\left(kR\right)^{2}+2kR$, where $R$ is the radius of the sphere circumscribing an antenna and $k$ is the free-space wave number. 
Kildal et al.~\cite{kildal_fundamental_2007,kildal_degrees_2017} improved the formula for directivity limit in the case of small-size antennas to $\mathcal D_{\rm lim}=\left(kR\right)^{2}+3$.
The above expressions for $\mathcal D_{\rm lim}$ concern the so-called {\em normal} gain \cite{Harrington1958} and rely on the hypothesis which postulates a {\em linear} relationship between the antenna size and the number of spherical harmonic modes, {$kR=\ell$}~\cite{Harrington1958} (see also the field degrees of freedom~\cite{bucci_degrees_1989}), that can be efficiently excited by the antenna.
The relation of the power radiated by the partial modes to the stored energy yields estimates on $Q$ and $\mathcal D$ \cite{hansen_fundamental_1981,geyi_physical_2003,collin_evaluation_1964}. 
However, the directivity is generally known to exceed these limits~\cite{bouwkamp_problem_1945,uzkov1946approach}. The antennas with $\mathcal D>\mathcal D_{\rm lim}$  are referred to as {\em superdirective}. 

Recent publications \cite{pigeon_miniature_2014,jonsson_methods_2017,pfeiffer_fundamental_2017,shahpari_fundamental_2018,jelinek_radiation_2018,chen_design_2019} show continuing interest in overcoming the physical limitations on antenna design. Most of them 
are related to emerging applications and to advances in computer-based approaches to antenna optimization~\cite{karlsson_efficiency_2013} as well as to numerical stability of theoretical treatments~\cite{Majic2020}. 
Since the superdirectivity of small antennas originates from their resonant behavior, it is accompanied by an ultranarrow-band response, which limits their applications. 
Nonetheless, the internet of things and wireless power transfer can benefit from such antennas in the radio frequency band~\cite{lin_electrically_2019}. 
In the optical band, high-index dielectric and plasmonic nanoparticles were used to enhance the direction-selective absorption and emission of nanoantennas~\cite{ziolkowski_using_2017}. 
A convex optimization based on the method of moments developed in Ref.~\cite{gustafsson_optimal_2013} was used to determine optimal surface currents which allowed both maximum gain and superdirective field patterns \cite{gustafsson_maximum_2019,gustafsson_tradeoff_2019} to be achieved. Although the proposed method is very general, it is unclear how to implement the optimal currents in antenna devices. 
Here, on using rigorous analytic treatment, we consider a spherical multilayer antenna with a dipole source excitation to demonstrate that particular superdirective designs with operation limits comparable to optimal current configurations~\cite{gustafsson_maximum_2019} are possible even in a relatively simple geometry of small {\em lossless} dielectric resonant antennas. 

\section{Theory}
\label{sec:optim}
We consider a lossless dielectric nonmagnetic concentric spherical multilayer antenna excited by an electric point dipole source, $\vp$, located on the $z$ axis.
Because a dipole emits predominantly in a perpendicular direction to its axis, the far-field radiation of a radially oriented dipole (i.e. parallel to the $z$ axis) is much more difficult to tailor by a nearby spherical antenna than for the tangential dipole orientation (i.e., perpendicular to the $z$ axis). For instance, for a radially oriented dipole outside the antenna, the far-field radiation part largely escapes to infinity without ever interacting with the antenna.
Out of two possible orthogonal dipole orientations, we thus focus on the {\em tangential} dipole orientation (see Supplemental Material Sec. I \cite{Gap_suppl}). 
The center of coordinates is at the sphere's origin. 
We denote the dipole position, $r_d$; radii of layers, $R_n$ (where $n=1,\dots, N$); homogeneous isotropic concentric domains, $\Omega_n$; refractive indices of spherical layer domains, $\eta_n \geq 1$; refractive index of a surrounding medium, $\eta_{N+1}=1$ (in the domain $\Omega_{N+1}$). We consider non-magnetic materials with the permeability of a vacuum. Dipole locations both inside and outside the spherical layers are allowed.

Maxwell's equations in each homogeneous layer can be solved via the vector spherical wave expansion of the electric field \cite[Eq. (8)]{moroz_recursive_2005}:
\begin{align}
{\bf E}_\gamma ({n, \bf r})=& \sum_L \left[
A_{\gamma L} (n) {\bf J}_{\gamma L}(k_n, {\bf r}) + B_{\gamma L}(n)
{\bf H}_{\gamma L}(k_n, {\bf r})\right],
\label{eq:E_H}
\end{align}
where $\vJ_{\gamma L}$ and $\vH_{\gamma L}$ are varieties of the vector multipoles $\vF_{\gamma L}$ (see Eq. (S1) in the Supplemental Material \cite{Gap_suppl}) which correspond to different linear combinations, $f_{\gamma \ell}$, of the spherical Bessel function in a given $n$-th shell \cite{moroz_recursive_2005}. 
Subscript $\gamma$ denotes the polarization: $\gamma=M$ for magnetic, or transverse electric (TE) polarization, $\gamma=E$ for electric, or transverse magnetic (TM) polarization.
The vector multipoles ${\bf F}_{M L}$ and ${\bf F}_{E L}$ reduce to $\mathcal{M}_{L}$ and $\mathcal{N}_{L}$ in the Stratton's notations \cite{Stratton_1941, Bohren1998} provided that $f_{\gamma \ell}$ reduces to one of $j_\ell$ or $h_\ell^{(1)}$. 
The index $L=(\ell,m)$ incorporates the orbital and magnetic quantum numbers.

The continuity of tangential field components determines the interface conditions allowing to relate the expansion coefficients $A_{\gamma L}$ and $B_{\gamma L}$ in adjacent neighboring regions $\Omega_n$ and $\Omega_{n+1}$ via $2 \times 2$ lowering $T_{\gamma L}^{-}$ or raising $T_{\gamma L}^{+}$ transfer matrices \cite{moroz_recursive_2005}:
\begin{equation}
\left(\begin{array}{c}
A_{\gamma L}(n+1)\\
B_{\gamma L}(n+1)
\end{array}\right)=T_{\gamma L}^{+}{(n)}\left(\begin{array}{c}
A_{\gamma L}{(n)}\\
B_{\gamma L}{(n)}
\end{array}\right),\label{eq:t-matrix1}
\end{equation}
\begin{equation}
\left(\begin{array}{c}
A_{\gamma L}(n)\\
B_{\gamma L}(n)
\end{array}\right)=T_{\gamma L}^{-}{(n)}\left(\begin{array}{c}
A_{\gamma L}{(n + 1)}\\
B_{\gamma L}{(n + 1)}
\end{array}\right).\label{eq:t-matrix2}
\end{equation}
The transfer matrix method of Ref. \cite{moroz_recursive_2005} enables one to calculate the emission of the dipole located either inside or in the vicinity of an arbitrary multilayered sphere. Two boundary conditions are required to unambiguously fix the solution. 
The so-called {\em regularity} boundary condition guarantees that the fields are not singular at the sphere core layer. Similarly to the case of Mie scattering from homogeneous spheres, this  requires that only $f_{\gamma \ell}\sim j_\ell$ are allowed in the core layer for $r<r_d$.
Unlike the case of Mie scattering, 
the other boundary condition depends on the position of the radiating dipole. 
For the dipole outside the multilayer sphere, the incoming dipole wave multipole
coefficients $A_{\gamma L}(N+1)$ have to be the coefficients $\alpha^d_{\gamma L}$
of the electric field of a radiating dipole source located at a position ${\bf r}_d$ of a 
homogeneous space \cite{chew_model_1976} [cf. Eq. (S10) in the Supplemental Material \cite{Gap_suppl}].
For the dipole inside the multilayer sphere,
there is no incoming wave arriving from the outside, hence $A_{\gamma L}(N+1) \equiv 0$.
However, in the layer where the radiating dipole is located, the field expansion becomes \cite{moroz_recursive_2005} 
\begin{equation}
\vE(\textbf{r})=
\left\{
\ba{ll}
\sum_{\gamma L}\left[\big(A_{\gamma L}+\alpha^d_{\gamma L}\big)\vJ_{\gamma L} + B_{\gamma L} \vH_{\gamma L}\right], & r< r_{d},
\\
\sum_{\gamma L}\left[A_{\gamma L} \vJ_{\gamma L} + \big(B_{\gamma L}+a^d_{\gamma L}\big) \vH_{\gamma L}\right], & r> r_{d},
\ea\right.
\label{eq:dipole}
\end{equation}
with the amplitudes having a {\em discontinuity} at $r=r_{d}$, because such is the dipole field (cf. Eqs. (S7)-(S10) in the Supplemental Material \cite{Gap_suppl}).
The quantities of crucial interest are the amplitudes $B_{\gamma L}{(N+1)}$ for $r>r_d$, which determine the radiating field escaping to infinity through the far-field amplitude matrix
${\bf F} (\theta,\varphi)$ (Eq. (S25) in the Supplemental Material \cite{Gap_suppl}).
What is very convenient in the case of a multilayered sphere is that the
$B_{\gamma L}{(N+1)}$'s can be represented in terms of $m$-{\em independent} linear combinations of the dipole field expansion coefficients $a^d_{\gamma L}$ and $\alpha^d_{\gamma L}$. The final solutions for $B_{\gamma L}{(N+1)}$ coefficients are strongly dependent on the position of the dipole source $r_d$. Four separate cases can be distinguished
\cite[Eqs. (60), (63), (65), (66)]{moroz_recursive_2005}:
\begin{itemize}

\item[(a)] For a dipole in a generic sphere shell different from the core and ambient:
\begin{equation}
B_{\gamma L}{(N+1)}= \frac{{\cal T}_{11;\gamma \ell}a^d_{\gamma L} + {\cal T}_{21;\gamma \ell} \alpha^d_{\gamma L}}
{{\cal M}_{22;\gamma \ell} {\cal T}_{11;\gamma \ell}-{\cal M}_{12;\gamma \ell}{\cal T}_{21;\gamma \ell}}
\nn
\end{equation}

\item[(b)] For a dipole in the sphere core:
\begin{equation}
B_{\gamma L}{(N+1)} = a^d_{\gamma L}/ {\cal M}_{22;\gamma \ell} (1)
\nn
\end{equation}

\item[(c)] For a dipole outside the sphere with $r<r_d$:
\begin{equation}
B_{\gamma L}{(N+1)} = \left[ {\cal T}_{21;\gamma \ell}(N+1)/
{\cal T}_{11;\gamma \ell}(N+1)\right] \alpha^d_{\gamma L}
\nn
\end{equation}

\item[(d)] for a dipole outside the sphere with $r>r_d$
\begin{equation}
B_{\gamma L}{(N+1)} = a^d_{\gamma L} +\left[ {\cal T}_{21;\gamma \ell}(N+1)/
 {\cal T}_{11;\gamma \ell}(N+1)\right] \alpha^d_{\gamma L},
\nn
\end{equation}

\end{itemize}
where ${\cal T}_{\gamma L} (n) = \prod_{x=1}^{n-1} T_{\gamma L}^+ (x)$ and ${\cal M}_{\gamma L}(n) = \prod_{x=n}^N T_{\gamma L}^-(x)$ are the ordered products of transfer matrices introduced formally by Eqs. (\ref{eq:t-matrix1}) and (\ref{eq:t-matrix2}).

Once the amplitudes $B_{\gamma L}{(N+1)}$ are known for \mbox{$r>r_d$},
the directivity in the direction specified by spherical angles $\left(\theta_0,\varphi_0\right)$ is determined by the formula
\begin{equation}
\mathcal D(\theta_0,\varphi_0)=\fr{4\pi r^2 |{\bf E} ({\bf r})|^2}{r^2 \oint |{\bf E} ({\bf r})|^2 d\Om},
\label{eq:directivity}
\end{equation}
where $|{\bf E} ({\bf r})|^2$ and $\oint |{\bf E} ({\bf r})|^2 d\Om$ can be expressed via coefficients $B_{\gamma L}(N+1)$ (see Supplemental Material Sec. I \cite{Gap_suppl}). Therefore, the directivity $\mathcal D\left(\theta_0,\varphi_0\right)$ depends on the geometrical and physical parameters of the problem $R_n$, $\eta_n$, $r_{d}$, the dipole polarization through the matrix elements of ${\cal T}_{\gamma L}$, ${\cal M}_{\gamma L}$ and dipole amplitudes $a_{\gamma L}^{d}$ and $\alpha_{\gamma L}^{d}$.

\section{Results and discussion}
\label{sec:results}
The best possible theoretical directivity is given by the $\delta(\theta,\varphi)$~function~\cite{Arslanagic2018}, where physical realization is a plane wave.
In the case of the $\delta$-function directivity, the expansion coefficients for the fields outside a sphere for $r>r_d$ have to be:
\begin{equation}
 B_{\gamma \ell 1}{(N+1)}\sim i^\ell \sqrt{\frac{2\ell+1}{4\pi}}
\label{D_coef}
\end{equation}
(cf. plane wave expansion in Eq. (4.37) of Ref. \cite{Bohren1998}).
Here $i$ denotes the imaginary unit.
The feasibility of this target, which imposes constraints on the geometry and material properties, is examined in the Supplemental Material Sec. I \cite{Gap_suppl}. It is shown there that the target can, in principle, be met only for the {\em tangential} dipole orientation. The coefficients Eq.~(\ref{D_coef}) lead to the Harrington directivity limit $\mathcal D_{\rm lim}=\ell_{\rm max}(\ell_{\rm max}+2)$ \cite{harrington_effect_1960} (cf. Eq. (S32)), where $\ell_{\rm max}$ is a cutoff on the summation over $\ell$.
Result Eq.~(\ref{D_coef}), which can be seen as a generalization of the results of Ref.~\cite{Arslanagic2018} for cylindrical geometry in two-dimensional space, is not bounded to a particular geometry and provides a recipe for an ultimate superdirective antenna design in three-dimensional space, useful for a range of applications. 

\subsection{Homogeneous sphere}

\begin{figure}[b!]
\begin{centering}
\includegraphics[width=8.6cm]{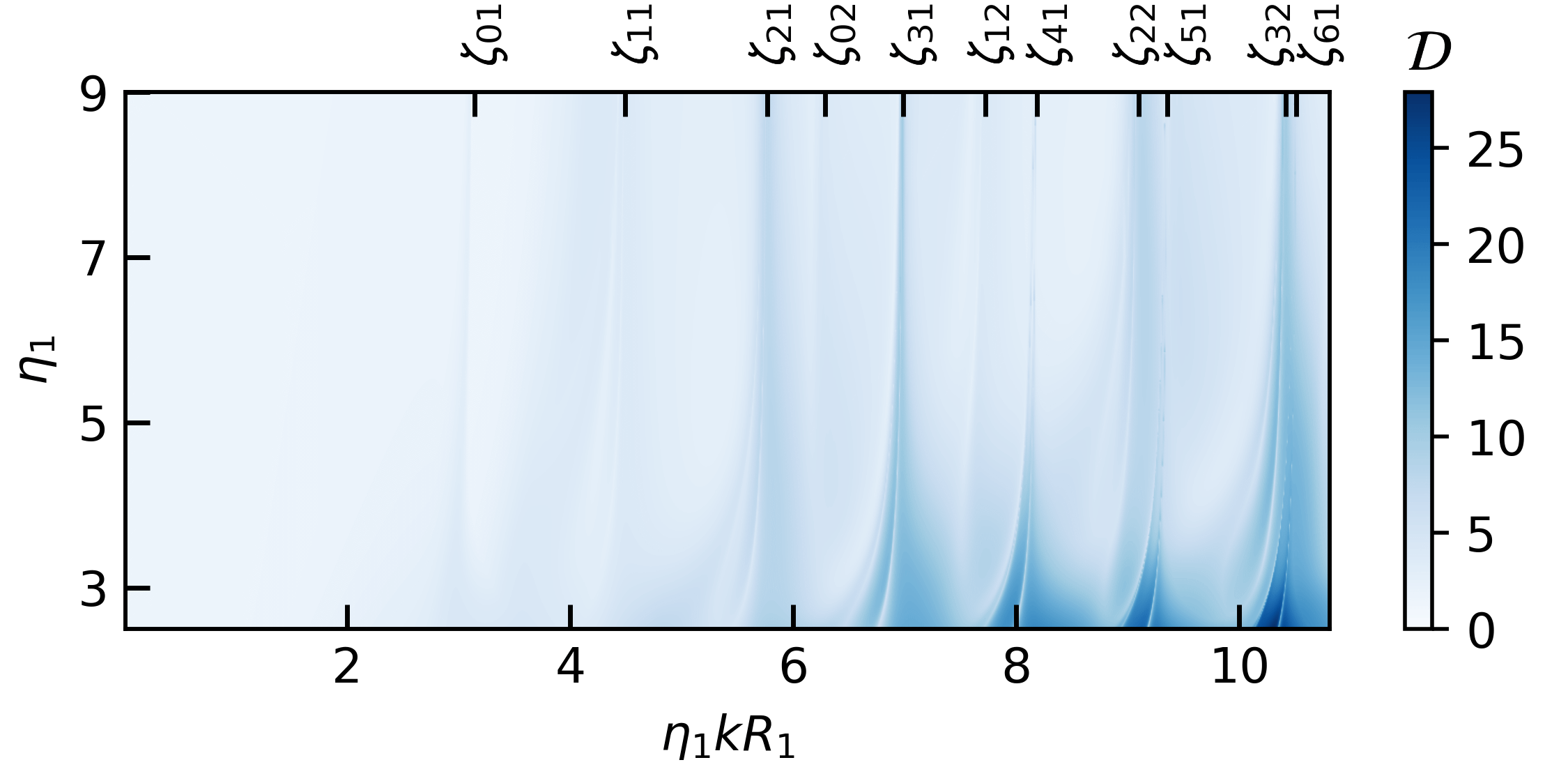}
\par\end{centering}
\caption{\label{fig:n_vs_nkR}
Optimized directivity $\mathcal D$ of a homogeneous spherical antenna excited by a point electric dipole source at its optimized position as a function of the size parameter $\eta_1kR_1$ and refractive index $\eta_1$.}
\end{figure}

\begin{figure}[t!]
\begin{centering}
\includegraphics[width=8.6cm]{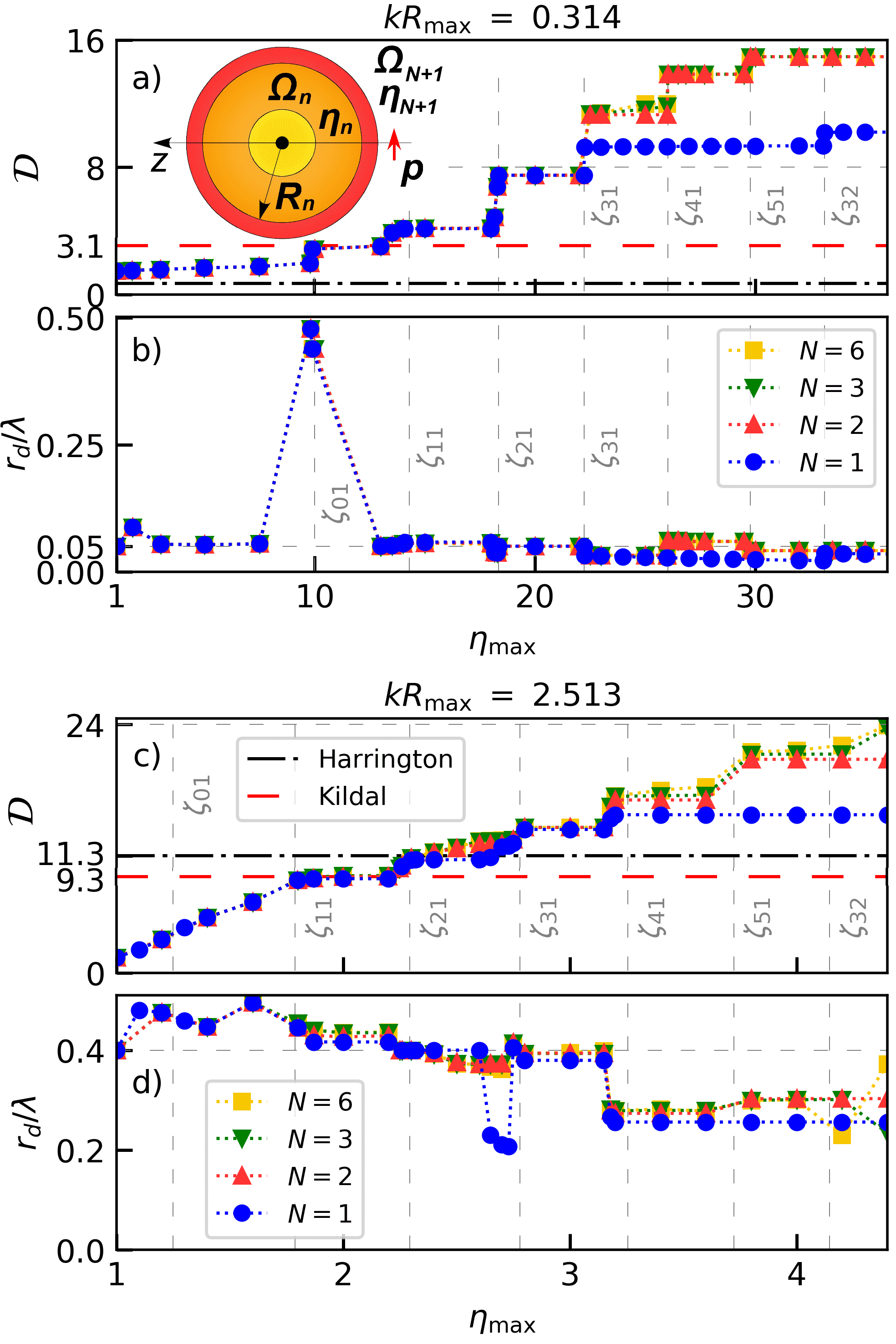}
\par\end{centering}
\caption{\label{fig:D_vs_n}
Maximum achievable directivity and optimum dipole position as a function of the upper bound, $\eta_{\rm max}$, on refractive index for antennas with a different number of layers $N$.
Sketch of the problem under the study is shown in inset in (a). The Harrington and Kildal limits, $\mathcal D_{\rm lim}$, are shown in (a) and (c) with black dash-dotted and red dashed lines, respectively.}
\end{figure}

A special case of a homogeneous spherical antenna represents an example worth careful examination.
Figure~\ref{fig:n_vs_nkR} shows optimized directivity as a function of the sphere's refractive index, $\eta_1$, and its size parameter, $\eta_1 k R_1$.
For each pair of $\{\eta_1 ; \eta_1 k R_1\}$, a dipole position, $r_d$, has been optimized to get maximum $\DD$ [see Figs.~\ref{fig:D_vs_n}(b) and ~\ref{fig:D_vs_n}(d)].
Rapid variations in directivity in Fig.~\ref{fig:n_vs_nkR} are associated with excitation of  TE$_{(\ell+1)ms}$ and TM$_{\ell ms}$ resonance modes in the dielectric sphere according to an approximate condition~\cite{Gastine1967,Lam1992,Balanis2012,gaponenko21},
\begin{equation}
    j_{\ell}(\zeta_{\ell s}) \simeq 0 \ ,
    \label{eq:multcond}
\end{equation}
where $\zeta_{\ell s}=\eta_1 k R_1$ is the $s$-th zero of the $\ell$-th order spherical Bessel functions of the first kind, $j_\ell$; $\eta_1$ and $R_1$ are refractive index and radius of a homogeneous spherical dielectric resonator.  
Table \ref{table:rescon} shows values of $\zeta_{\ell s}$ for small $s$ and $\ell$. 
These resonances exhibit themselves by the leakage of trapped electromagnetic field out of high refractive index dielectric scatterer into the environment, leading to resonances of scattered field (interaction between the polarization energy stored in the dielectric and the energy stored in the magnetic field \cite{Forestiere2019}).

\begin{table}
\centering
\caption{\label{table:rescon}
Zeros $\zeta_{\ell s}$ of the spherical Bessel function of the first kind $j_{\ell}(\zeta_{\ell s})=0$. The subscript $s$ denotes the ordinal number of the zero of the $\ell$-th order spherical Bessel function.}
\begin{ruledtabular}
\begin{tabular}{ c | c c c }
   $ $ & $s=1$     & $s=2$     & $s=3$   \\ [0.5ex]
 \hline
 $\ell=0$ & $3.14159$ & $6.28319$ & $9.42478$\\
 $\ell=1$ & $4.49341$ & $7.72525$ & $10.9041$\\
 $\ell=2$ & $5.76346$ & $9.09501$ & $12.3229$\\
 $\ell=3$ & $6.98793$ & $10.4171$ & $13.698$ \\
 $\ell=4$ & $8.18256$ & $11.7049$ & $15.0397$\\
 $\ell=5$ & $9.35581$ & $12.9665$ & $16.3547$\\
 $\ell=6$ & $10.5128$ & $14.2074$ & $17.648$ \\
\end{tabular}
\end{ruledtabular}
\end{table}

\begin{figure}[t!]
\begin{centering}
\includegraphics[width=8.6cm]{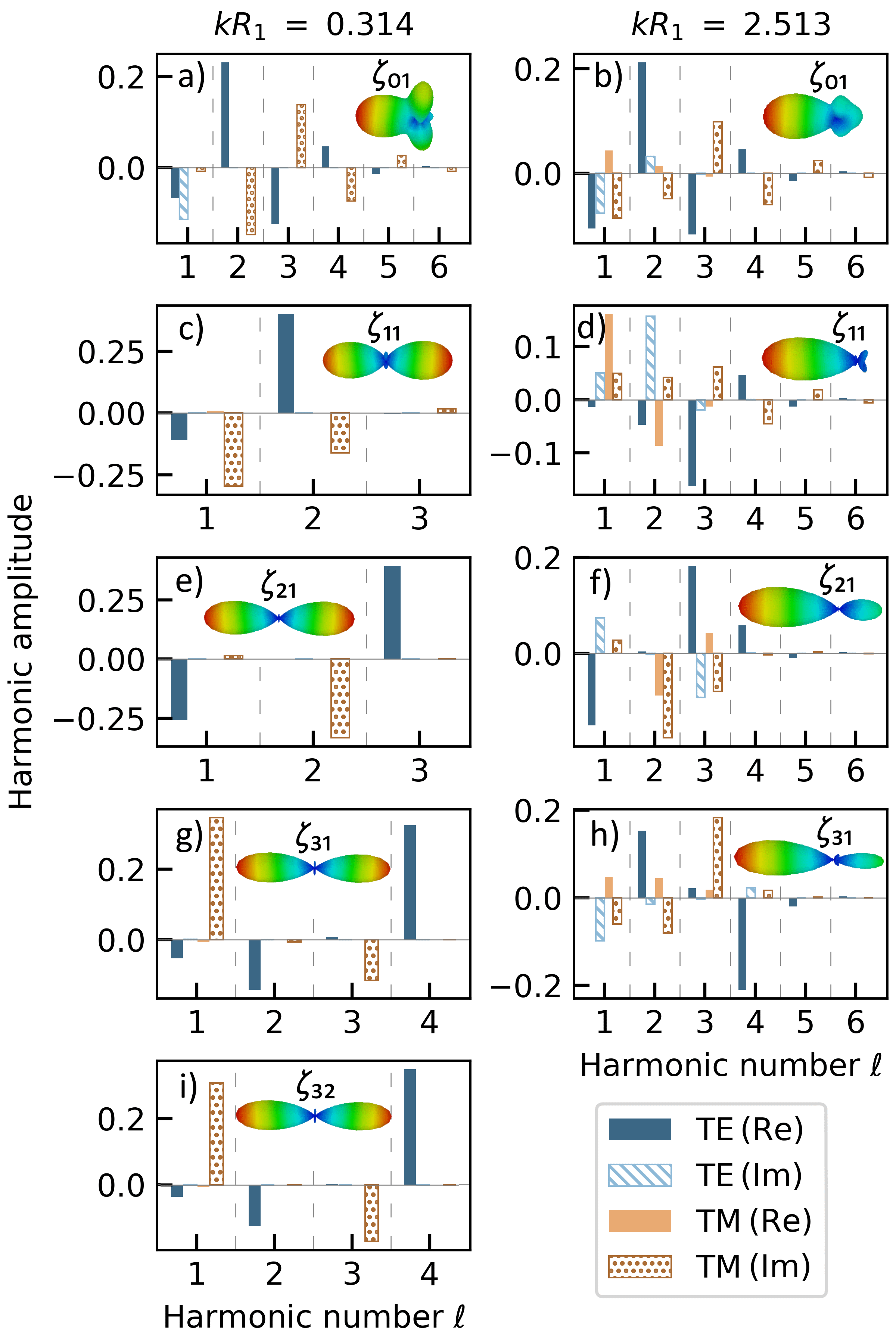}
\par\end{centering}
\caption{\label{fig:harmonics}Normalized harmonic amplitudes at points corresponding to the directivity jumps in Figure \ref{fig:D_vs_n} for homogeneous spheres with $kR_N=0.314$ and $kR_N=2.513$. The amplitudes are normalized to the sum of the absolute values of the amplitudes for all excited harmonics. Each subplot depicts the corresponding resonance conditions and shows the resulting far-field diagram calculated via full-wave numerical solver CST Studio Suite$^{\text{®}}$ 2019~\cite{CST}. Directivity extracted from far-field diagrams agrees with our theoretical predictions up to 1\% accuracy.}
\end{figure}

Figure~\ref{fig:D_vs_n} shows the dependence of optimized directivity on the refractive index, $\eta_{\rm max}$ for $kR_N \leq 0.314$ and $2.513$.
According to our numerical results, the maximum directivity for a homogeneous sphere is achieved at $\zeta_{3s}$ resonances. 
In general, $\zeta_{\ell s}$ resonances are mainly occur due to the simultaneous excitation of TE$_{(\ell+1)ms}$ and TM$_{\ell ms}$ modes ~\cite{gaponenko21}, which is shown in the harmonic expansion plots in Figure~\ref{fig:harmonics}. 
The accurate resonance conditions for these modes are different and the maximum directivity is achieved if all excited modes overlap with optimally tuned amplitudes and phases. 
The higher the order of overlapping modes, the harder their excitation, since the homogeneous sphere does not have additional degrees of freedom that would allow for simultaneous resonant tuning of multiple modes.
This is why the value of optimized directivity does not increase for $\zeta_{\ell s}$ resonances with $\ell>3$.

The position of the exciting electric dipole also affects the order, $\ell$, of harmonics excited in the antenna, see Fig.~\ref{fig:harmonics}.
If the dipole source is located at the center of a sphere, only $\ell=1$ is prominent, whereas higher harmonics are induced if the dipole is moved away from the center of the sphere. 
For example, only $\ell=1$ and $\ell=2$ harmonics are pronounced for the dipole at position $r_{d} = 0.05\lambda$, while $\ell=1, \dots, 6$ are prominent for $r_{d} = 0.4\lambda$, see Figs.~\ref{fig:D_vs_n}(b) and ~\ref{fig:harmonics}.
We notice that $\zeta_{01}$ resonance is unique, since in this case the electric dipole source excites only the TE$_{101}$ mode (magnetic dipole mode) of the sphere. In order to achieve the highest directivity in the forward direction, electric dipole source and excited magnetic dipole mode must have $\pi$ phase difference with an optimal dipole position $r_d\rightarrow\lambda/2$, see Fig.~\ref{fig:D_vs_n}(b) and Ref.~\cite{gaponenko21}.
Figure~\ref{fig:D_vs_n}(a) makes it clear that a single point electric dipole is not enough to excite resonances $\zeta_{\ell s}$ of homogeneous sphere with $\ell>3$.

Finally, for electrically small antennas with $kR_1\rightarrow 0$, we observe superdirective behavior at $\zeta_{3s}$ resonances. 
With increasing $s$, the directivity of the antenna continues to increase slightly: $\mathcal D>11.2$ for the resonances with $s>14$ (see Table~S1 in the Supplemental Material \cite{Gap_suppl}), while the optimal dipole position moves slightly toward the surface of a sphere.

\subsection{Multilayer sphere}

In what follows, we shall use Eq.~\eqref{eq:directivity} for direct evaluation of the directivity in the optimization problem for determining maximum directivity $\mathcal D\left(\theta_0,\varphi_0\right)$ for a given number of layers $N$.
We have optimized directivity using the following constraints:
\begin{enumerate}

\item Setting the value of the maximum allowable refractive index $\eta_{\rm max}$ of any layer of a sphere and setting the value of $kR_{\rm max}$ (see Fig.~\ref{fig:D_vs_n}). 
The refractive indices of layers, $\eta_n$, and their sizes, $R_n$, are continuously varied within $1\leq \eta_n \leq \eta_{\rm max}$ and $0< R_n \leq R_{\rm max}$ ranges;

\item Setting the maximum allowable sphere size parameter $kR_{\rm max}$ for two different constraints on the refractive indices of layers:

\begin{itemize}

\item Setting the value of the maximum allowable refractive index $\eta_{\rm max}$ and allowing the refractive indices of layers to be continuously varied in $1\leq \eta_n\leq \eta_{\rm max}$ range [see Fig.~\ref{fig:D_vs_size}(a)];

\item Setting the discrete set of allowable refractive indices \mbox{$\eta_n\in\left\{ \eta^{(1)},\eta^{(2)},\dots\right\}$}, where \mbox{$1 \leq \eta^{(\cdot)}\leq\eta_{\rm max}$} [see Fig.~\ref{fig:D_vs_size}(b)].
\end{itemize}

\end{enumerate}
 
The latter strategy is the closest to possible practical implementations, since the set of materials available for fabrication is essentially limited (e.g., high-index ceramics in the radio-frequency band \cite{dielectrics}, or high-index dielectrics in the optical band \cite{Baranov2017}), and it is instructive to compare optimization results in these two cases.
In any optimization strategy the position of the dipole, $r_d$, is not limited.

\begin{figure}[t!]
\begin{centering}
\includegraphics[width=8.6cm]{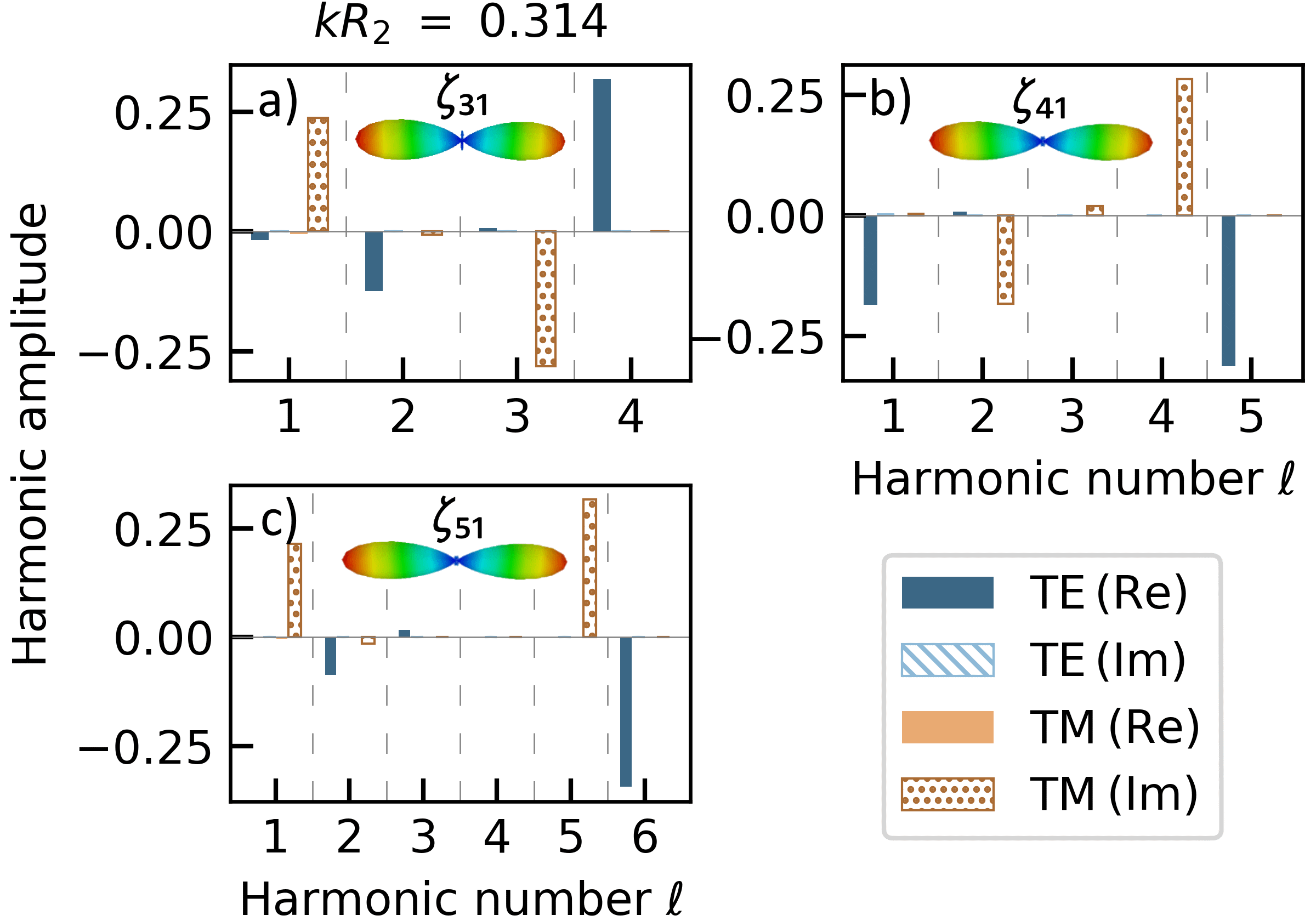}
\par\end{centering}
\caption{\label{fig:harmonics2}
Same as in Fig.~\ref{fig:harmonics}, but only for two-layer (core-shell) sphere with $kR_N=0.314$.}
\end{figure}

Since the convexity of Eq. (\ref{eq:directivity}) with respect to the optimization parameters is difficult to establish or check, we used a stochastic JADE algorithm \cite{Zhang2009} to maximize the directivity. This algorithm is an improved version of the differential evolution, which is effective for global optimization of functions with a large number of local peaks \cite{Zhang2009, Yang2014} and, in particular, for problems associated with electromagnetic scattering \cite{ladutenko_reduction_2014}.
For a convergence of the JADE optimization algorithm, one should select appropriate parameters (population size, number of generations, crossover probability, etc), which strongly depend on the $Q$ factor of the excited resonances: the higher the $Q$ factor, the larger value of these parameters is required. 
Moreover, a numerical implementation of the theoretical treatment presented above requires an accurate truncation of infinite sums to a limited number of terms. 
As discussed for spherical near-field antenna measurements~\cite{Standards2012}, the recommended number is $\ell_{\rm max}=kR_{\rm max}+x_1$, where $kR_{\rm max}$ is an integer closest to the wave number $k$ times the radius $R_{\rm max}=\max\{r_{d};R_{N}\}$, and $x_1$ is an integer which depends on the position of the source and desired accuracy, wherein $x_1=10$ is sufficient for most practical cases \cite{Hansen1988}.
Nonetheless, the relation linking the size parameter with $\ell_{\rm max}$ is still questionable for electrically small resonant antennas.

In the lossless case, there is a limit on the directivity for electrically small antennas which depends on the number of layers. Multilayer scatterers can provide higher directivity as shown in Fig.~\ref{fig:D_vs_n} and have more complex resonance conditions, which can be derived from expression Eq.~(\ref{eq:directivity}).
The small multilayer sphere with $\eta_{\rm max}kR_N < \zeta_{31}$ usually converges to a homogeneous case.
For $\eta_{\rm max}kR_N \geq \zeta_{31}$, directivity optimization results are different, since the additional internal layers can provide more efficient control over the field distribution and over the interference between different multipoles without changing the external size of the antenna. 
The stored energy of the lower order modes is concentrated closer to the center. Therefore, the modes are more sensitive to changes in the internal structure of the sphere
and the use of a multilayer structure of a sphere makes it possible to efficiently excite higher-order resonances.
Harmonic expansions for several resonances of the two-layer sphere are shown in Fig.~\ref{fig:harmonics2} (see Supplemental Material Sec. IV \cite{Gap_suppl} for respective data).
In the parameter region defined by $\eta_{\rm max}kR_N > \zeta_{51}$, it becomes difficult to obtain the global maximum numerically using stochastic optimization procedure for multilayer spheres due to a large number of narrow directivity peaks originating from an interplay between high-$Q$ ($>10^{10}$) TE and TM resonances. 
In this case, stochastic optimization (i) is computationally expensive, (ii) not necessarily converges to a global maximum, and (iii) requires a large number of initial populations and generations. 
There are various physical reasons which limit a practically achievable $Q$ factor of antennas: surface roughness~\cite{Farias1994,Li2004}, material imperfections, the presence of other objects in the immediate vicinity of the antenna, temperature of the environment, and challenging fabrication.

Figure~\ref{fig:D_vs_size} demonstrates the dependence of the optimized directivity $\mathcal D$ on the antenna size parameter $0.3 \leq kR_N \leq 12$ for different limitations on the refractive index of a sphere $\eta_{\rm max}$ (taking into account high-index ceramics in the microwave band).
\begin{figure}[t!]
\begin{centering}
\includegraphics[width=8.6cm]{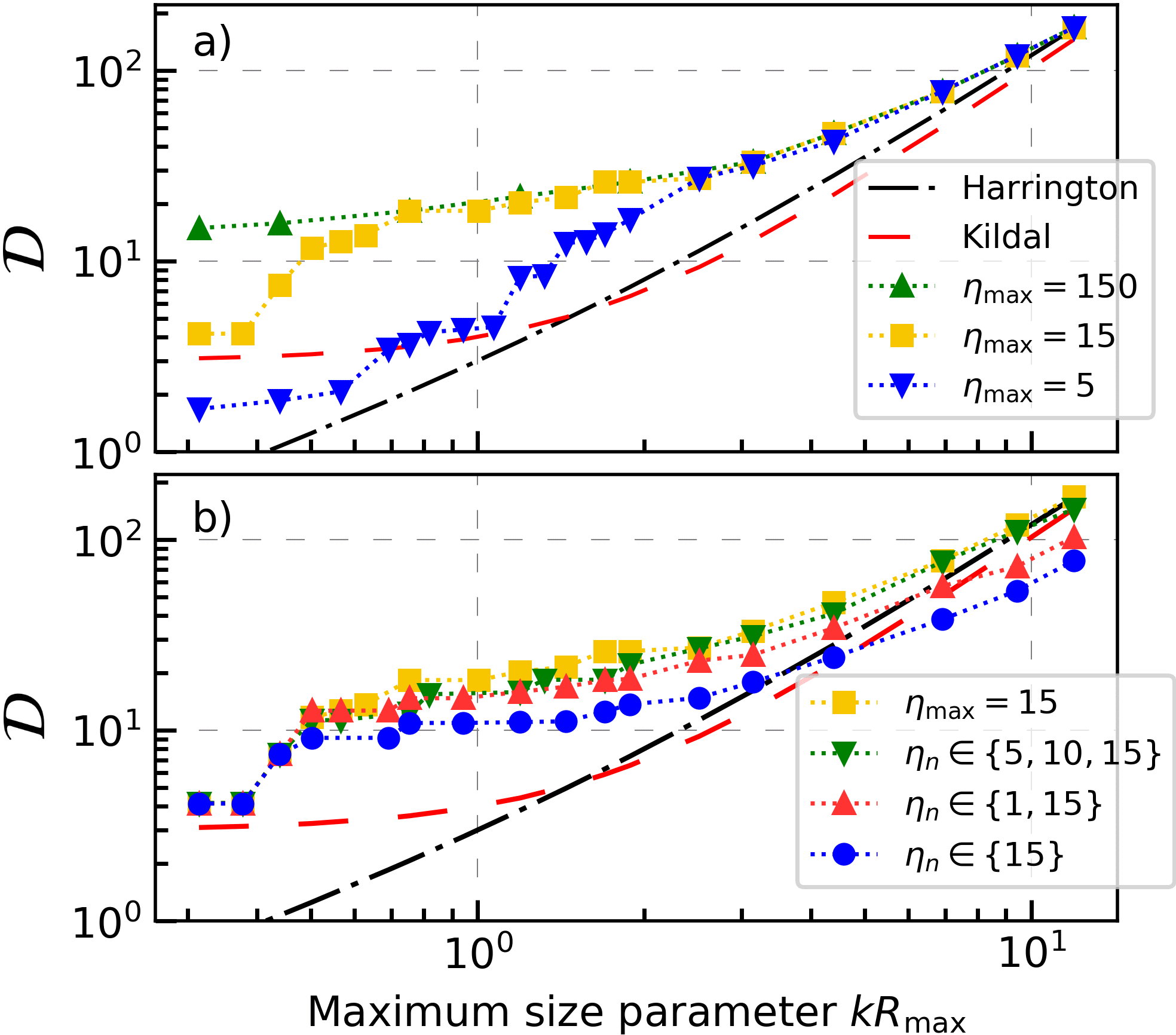}
\par\end{centering}
\caption{\label{fig:D_vs_size}Dependence of the optimized directivity on the maximum size parameter $kR_{\rm max}$ for a spherical three-layer antenna with a different limitation on a refractive index: (a) continuously varying, (b) discrete set.}
\end{figure}
Directivity behaves as a step function where the peaks correspond to the interference of the electric dipole with the various induced multipoles of the sphere (either magnetic or electric).
Figure~\ref{fig:D_vs_size}(a) shows that the use of dielectric materials with $\eta_{\rm max} = 15$ allows us to reach the theoretically predicted gain limit for electrically small copper antennas at $1$GHz (cf. curve with $10^{-2}\Omega/\square$ surface resistivity in Fig. 1 of Ref.~\cite{gustafsson_maximum_2019}). Higher refractive index, $\eta_{\rm max}$, or higher applied frequency, in principle, allow us to overcome this limit, which makes high-index ceramics even more attractive.
Also, the optimal directivity is almost independent on the formulation of the optimization problem for sufficiently large sets of available refractive indices as demonstrated in \mbox{Fig.~\ref{fig:D_vs_size}(b)}. $\mathcal D$ depends not only on the number of terms in a set but also on the available refractive indices.

An example of optimization for a homogeneous sphere is presented in the Supplemental Material Sec. II \cite{Gap_suppl}. Several optimized designs of homogeneous and multilayer spherical superdirective antennas are presented in the Supplemental Material Sec. III \cite{Gap_suppl}. The resulting designs for $kR_N\rightarrow 0$ are generally forward-backward (as a result of constructive interference of the TE$_{(\ell+1)ms}$ and TM$_{\ell ms}$ modes in forward and backward directions), while as $kR_N$ increases, the {\em forward} direction becomes dominant due to the interference between different modes (see Fig. \ref{fig:D_vs_n}). 
Occasionally, more complex shapes can be obtained, because optimization is 
performed in the specified direction $\theta=0^{\circ}$ and does not control 
the field distribution in other directions.

\section{Conclusion}
\label{sec:summary}
Knowing an ideal upper limit on the antenna directivity does not answer the question if the limit can be physically realized. The knowledge of an ideal current configuration on a spherical surface is not of much help if there is no practical way of realizing it. It may well turn out that exotic, i.e. not achievable in practice, material properties will be required to produce ideal current configurations.
For a cylindrical antenna driven by a line source, saturating the directivity bounds for a given number $N$ of layers was already a tedious and difficult task requiring excessive optimization effort and a juxtaposition of positive and negative (i.e., metallic) material regions~\cite{Arslanagic2018}.

Our contribution can be seen in providing explicit engineering recipes for experimentally feasible designs of optimal directivity for a given number $N$ of shells of compact {\em spherical} multilayer antennas while employing only purely {\em dielectric} materials. 
The absence of any metal component makes our designs interesting for space applications.
Surprisingly, in spite of limiting ourselves to purely dielectric materials, we have shown that the classical limit for directivity of electrically small resonant antennas can be overcome. 
We presented (i) a number of superdirective antenna designs and (ii) 
a stochastic optimization-based recipe to construct new superdirective antennas 
of a given size and materials. We have also provided a rigorous derivation of 
the maximum directivity, $\mathcal D$, for the ultimate superdirective antenna design. 
Our performance analysis revealed that, as the size of antenna decreases, 
it is essential to use high refractive index materials to obtain 
a superdirective antenna radiation pattern.
An important property of the proposed designs is their geometrical simplicity: 
a multilayer sphere excited by a point dipole.
It is anticipated that multiple dipole feeds 
will be capable of communicating with multiple receiver devices simultaneously 
using highly directive beams without any mechanical rearrangement.
We hope that the results presented here could be applied in the development 
of subwavelength resonant dielectric antennas for the internet of things and other wireless applications.
Our results can be easily adapted for other cases, such as lossy metal materials, which are of significant interest~\cite{Arslanagic2018} for fluorescence collection and quantum optics, and are the subject of further study.
Last but not the least, a duality between electric and magnetic dipole sources enables one to perform an alternative study for a magnetic dipole driven antenna.



%


\end{document}



\title{Supplementary Material\\
Details of the Analytical and Numerical Results Reported in the Article: \\``Harnessing superdirectivity in dielectric spherical multilayer antennas''}

\author{Roman Gaponenko,$^{1}$ Alexander Moroz,$^2$ Ilia L. Rasskazov,$^{3}$ Konstantin Ladutenko,$^1$ Alexey Shcherbakov,$^{1}$ and Pavel Belov$^1$}
\affiliation{%
 $^1$ School of Physics and Engineering, ITMO University, 197101, Saint-Petersburg, Russia \\ $^2$ Wave-scattering.com \\ $^3$ The Institute of Optics, University of Rochester, Rochester, New York 14627, United States \\
* Corresponding authors: irasskaz@ur.rochester.edu; alexey.shcherbakov@metalab.ifmo.ru
}%

\maketitle

\clearpage
\setcounter{equation}{0}
\setcounter{figure}{0}
\setcounter{table}{0}
\renewcommand{\theequation}{S\arabic{equation}}
\renewcommand{\thefigure}{S\arabic{figure}}
\renewcommand{\thetable}{S\arabic{table}}

\section{Theoretical part}
\label{ap:A}

\subsection{Special values of the vector spherical harmonics and
dipole field expansion coefficients on the $z$-axis}
Normalized transverse vector multipoles in Eq.~(1) are defined as
\begin{eqnarray}
{\bf F}_{ML}(k_n,{\bf r}) &= & f_{ML} (k_n r) {\bf Y}^{(m)}_L({\bf r}),
\nonumber\\
{\bf F}_{EL}(k_n,{\bf r}) &=&
\frac{1}{k_n r}\left\{\sqrt{\ell(\ell+1)}f_{EL} (k_n r){\bf Y}^{(o)}_L({\bf r}) 
\right.
\nn\\
&& \left. 
+
\big(r f_{EL}(k_n r)\big)'\,{\bf Y}^{(e)}_L({\bf r})\right\},
\label{fvmultip}
\end{eqnarray}
where $f_{\gamma L}$ is an arbitrary linear combination of spherical
Bessel functions (${\bf F}_{\gm L} \equiv {\bf J}_{\gm L}$ for $f_{\gm L} = j_\ell$ and ${\bf F}_{\gm L} \equiv {\bf H}_{\gm L}$ for $f_{\gm L} = h^{(1)}_{\ell}$), 
${\bf Y}^{(a)}_L$ are vector spherical harmonics, 
and $L$ is a composite angular momentum index,
$L=(\ell,m)$, where the respective $\ell\ge 1$ and $m$ label 
the orbital and magnetic angular numbers,
and prime denotes the radial derivative, ${\rm d}/{\rm d}r$. 
In what follows it is expedient to introduce scalar angular functions,
\begin{eqnarray}
\pi_{m\ell}(\theta)=\frac{m}{\sin\theta}\, d_{0m}^\ell(\theta), \quad
\tau_{m\ell}(\theta)=\frac{\rm d}{{\rm d}\theta}\, d_{0m}^\ell (\theta),
\lb{pitau}
\end{eqnarray}
which are defined in terms of the Wigner $d$-functions $d_{0m}^\ell$ \ct{Edmonds1960} and
which can all be generated by stable recurrences. Then with the orthonormal spherical coordinate basis vectors $\ve_r, \ve_\theta, \ve_\phi$:
\bea
\vY^{(m)}_L &=& (-1)^m i d_\ell \left(i \ve_\theta \,\pi_{m\ell} - \ve_\phi  \,
\tau_{m\ell} \right) e^{im\vphi}
\nn\\
&=& 
i (-1)^m d_\ell {\bf C}_L(\theta) e^{im\varphi},
\nn\\
\vY^{(e)}_L &=& (-1)^m i d_\ell \left( \ve_\theta\,\tau_{m\ell} + i \ve_\phi \,
\pi_{m\ell} \right) e^{im\vphi}
\nn\\
&=&
i (-1)^m d_\ell {\bf B}_L(\theta) e^{im\varphi} ,
\nn\\
\vY^{(o)}_L &=& i \ve_r Y_L=i \gm_L'\, P_\ell^m (\cos\theta)\, e^{im\vphi} \ve_r
\nn\\
&=& (-1)^m i \sqrt{\ell(\ell+1)}\, d_\ell \, d_{0m}^\ell (\theta)\, e^{im\vphi} \ve_r
\nn\\
&=& (-1)^m i \sqrt{\ell(\ell+1)}\, d_\ell \, {\bf P}_L (\theta)\, e^{im\vphi},
\lb{vectsphec}
\eea
where $Y_L$ with $\ell\ge 1$ are the usual orthonormal scalar spherical harmonics in the Condon-Shortley convention
defined in terms of associate Legendre functions $P_\ell^m (\cos\theta)$ as, for instance, by Jackson \cite{Jackson1999}, and
the numerical constant $\gamma_L'$ of Ref. \cite{Tsang1985} and numerical constant $d_\ell$ of Ref. \cite{Mishchenko1991}, are
\begin{equation}
d_\ell=\left[\frac{2\ell+1}{4\pi \ell(\ell+1)}\right]^{1/2},
\quad
\gamma_L' = \sqrt{ \frac{(2\ell+1)(\ell-m)!}{4\pi(\ell+m)!} }\cdot
\lb{dldf}
\end{equation}

$\vC$, $\vB$, and $\vP$ can be expressed as \ct{Mishchenko1991} 
\bea
\vC_L(\theta) &=&
\ve_\theta \fr{im}{\sin\theta}
d_{0m}^\ell (\theta) - \ve_\phi \frac{{\rm d} d_{0m}^\ell (\theta)}{{\rm d}\theta},
\nn\\
\vB_L (\theta) &=&
\ve_\theta \frac{{\rm d} d_{0m}^\ell (\theta)}{{\rm d}\theta} +\ve_\phi \fr{im}{\sin\theta}
d_{0m}^\ell (\theta) ={\bf r}_0\times\vC_L,
\nn\\
 \vP_L(\theta) &=& \fr{{\bf r}}{r}\, d_{0m}^\ell (\theta).
 \lb{mishbcp}
\eea
Eq. (4.2.3) of \ct{Edmonds1960} for the special case $\bt=\theta=0$ and $m'=0$ yields
\bg
d_{0m}^\ell(0) = (-1)^{m} \dt_{0m} = \dt_{m0}.
\lb{ed4.2.3}
\eg
It should be remembered though that Edmonds $d_{0m}^\ell$ are related to our
up to prefactor of $(-1)^m$. However, because of $\dt_{0m}$, the latter factor
does not play any role in the above identities.

The electric field at ${\bf r}$ due to an electric dipole ${\bf p}$ at ${\bf r}_d$
radiating at frequency $\omega$ in a medium with $\varepsilon$, $\mu$ is given for $r>r_d$ by \cite{Chew1987}
\begin{eqnarray}
{\bf E}_d ({\bf r}) & = & \sum_L \left[
 a^d_{ML}{\bf H}_{ML}(k,{\bf r})+ a^d_{EL}{\bf H}_{EL}(k,{\bf r}) \right],
\nonumber\\
{\bf H}_d ({\bf r}) & = &-i\sqrt{\frac{\varepsilon}{\mu}}\, \sum_L \left[ 
a^d_{ML} {\bf H}_{EL}(k,{\bf r})+ a^d_{EL} {\bf H}_{ML}(k,{\bf r})\right],\nonumber\\
\label{dipemf}
\end{eqnarray}
where $\ell\ge 1$ and \cite[Eq. (44)]{moroz_recursive_2005}
\begin{eqnarray}
a^d_{ML} &=& 
4\pi i (k^3/\varepsilon) \, {\bf p}\cdot {\bf J}_{ML}^*(k,{\bf r}_d),\nonumber\\
a^d_{EL} &=& 4\pi i (k^3/\eta) \, \sqrt{\frac{\mu}{\varepsilon}}
 \, {\bf p}\cdot {\bf J}_{EL}^*(k,{\bf r}_d)
\nonumber\\
 &=& 4\pi i (k^3/\varepsilon) \, {\bf p}\cdot {\bf J}_{EL}^*(k,{\bf r}_d),
\label{as}
\end{eqnarray}
$k$ being the wave vector in the medium where the dipole is 
embedded. For $r<r_d$, one interchanges $j_\ell$ and $h_\ell^{(1)}$ and has
\begin{eqnarray}
{\bf E}_d ({\bf r}) & = & \sum_L \left[
 \alpha^d_{ML} {\bf J}_{ML}(k,{\bf r})+ \alpha^d_{EL}{\bf J}_{EL}(k,{\bf r}) \right],
\nonumber\\
{\bf H}_d ({\bf r}) & = &-i\sqrt{\frac{\varepsilon}{\mu}}\, \sum_L \left[ 
\alpha^d_{ML} {\bf J}_{EL}(k,{\bf r})+ \alpha^d_{EL} {\bf J}_{ML}(k,{\bf r})\right],\nonumber\\
\label{diphmf}
\end{eqnarray}
where \cite[Eq. (46)]{moroz_recursive_2005}
\begin{eqnarray}
\alpha^d_{ML} &=& 
4\pi i (k^3/\varepsilon) \, {\bf p}\cdot {\bf H}_{ML}^*(k,{\bf r}_d),
\nonumber\\
\alpha^d_{EL} &=& 4\pi i (k^3/n) \, \sqrt{\frac{\mu}{\varepsilon}}
\, {\bf p}\cdot {\bf H}_{EL}^*(k,{\bf r}_d)
\nonumber\\
 &=& 4\pi i (k^3/\varepsilon) \, {\bf p}\cdot{\bf H}_{EL}^*(k,{\bf r}_d).
\label{alphas}
\end{eqnarray} 
To avoid confusion, star here and in Eq. (\ref{as}) above denotes the complex conjugation 
which only applies to the vector spherical harmonics and not to the spherical 
Bessel functions \cite{Chew1987,Chew1988}.

In what follows, the above formulas (\ref{vectsphec}) for vector spherical harmonics 
will be used in reducing the electric dipole fields (\ref{dipemf}) and (\ref{diphmf})
for the special case when the dipole is located on the $z$-axis.

\subsubsection{$\theta=0$}

According to Eqs. (4.1-4) of \ct{Mishchenko1991} (cited therein as
formulae of D. A. Varshalovich, A. N. Moskalev, and
V. K. Khersonskij, {\em Quantum Theory of Angular Momentum}
(Nauka, Leningrad, 1975))
\begin{eqnarray}
\pi_{m\ell}(0) &=& \frac{\delta_{m,\pm 1}}{2}\, \sqrt{\ell(\ell+1)},
\nonumber\\
\tau_{m\ell}(0) &=& \frac{m \delta_{m,\pm 1}}{2}\,
\sqrt{\ell(\ell+1)}.
\lb{vmk1}
\end{eqnarray}
On substituting the special values
(\rf{ed4.2.3}) of the Wigner functions and (\rf{vmk1}) in (\rf{mishbcp}), one finds
\bea
\vC_L(0) &=&
\fr{\sqrt{\ell(\ell+1)}}{2}\, (i\ve_\theta -m \ve_\phi) \dt_{m,\pm 1},
\nn\\
\vB_L (0) &=& \fr{\sqrt{\ell(\ell+1)}}{2}\, (m \ve_\theta +i \ve_\phi) \dt_{m,\pm 1},
\nn\\
 \vP_L(0) &=& \hat{\vz} \, \dt_{m0}.
\nn 
\eea
The values of the vector spherical harmonics in (\rf{vectsphec}) for ${\bf r}$ on the positive $z$-axis
follow directly on multiplying the above results by the factor $i(-1)^m d_\ell$,
and additionally by $\sqrt{\ell(\ell+1)}$ for $\vY^{(o)}_L$, resulting in

\bea
\vY^{(m)}_L (\hat{\vz})
 &=& - i \sqrt{\fr{2\ell+1}{4\pi}}\,\fr{\dt_{m,\pm 1}}{2}\,
\left(i \ve_\theta - m \ve_\phi \right) e^{im\vphi}
\nn\\
&=&
 \fr{\dt_{m,\pm 1}}{2}\, \sqrt{\fr{2\ell+1}{4\pi}}\,
\left(\ve_\theta +i m \ve_\phi \right) e^{im\vphi},
\nn\\
\vY^{(e)}_L (\hat{\vz})
 &=& - i \sqrt{\fr{2\ell+1}{4\pi}}\,\fr{\dt_{m,\pm 1}}{2}\,
\left(m \ve_\theta + i \ve_\phi \right) e^{im\vphi}
\nn\\
&=&
\fr{\dt_{m,\pm 1}}{2}\, \sqrt{\fr{2\ell+1}{4\pi}}\,
\left(- im \ve_\theta + \ve_\phi \right) e^{im\vphi},
\nn\\
\vY^{(o)}_L (\hat{\vz})
 &=& i \dt_{m0}\, \sqrt{\fr{2\ell+1}{4\pi}}\, \hat{\vz}.
\lb{vctsphsv}
\eea
One can easily verify that general relations
$\vY^{(c)*}_L=-(-1)^m \vY^{(c)}_{\ell,-m}$ hold for the special values.
The above values in spherical coordinates can be translated into those in Cartesian
coordinates on using elementary relations
\bea
\ve_z &= \cos\theta\, \ve_r -\sin \theta\, \ve_\theta,
\quad
\ve_\rho & = \sin\theta\, \ve_r +\cos \theta\, \ve_\theta,
\nn\\
\ve_x &= \cos \varphi\, \ve_\rho - \sin \varphi\, \ve_\varphi,
\quad
\ve_y &= \sin\varphi\, \ve_\rho + \cos \varphi\, \ve_\varphi.
\nn
\eea

For a dipole located on the $z$-axis in the $\phi=0$ plane we have two possible
orientations: along either (i) $\ve_\theta$ (i.e. tangential) or (ii) $\ve_r$ (i.e. radial). In the former case,
 on making use of Eq. (3) of Ref. \cite{moroz_recursive_2005} 
 and (\ref{vctsphsv}),
the dipole field expansion coefficients $a^d_{\gm L}$ of 
Eq. (\ref{as})
 become
\begin{eqnarray}
a^d_{ML} &=& i \dt_{m,\pm 1} \left[\sqrt{(2\ell+1)\pi}\, (k_d^3/\varepsilon_d)\, p_\theta \,\right]\, j_\ell(k_dr_d),
\nonumber\\
a^d_{EL} &=& m \dt_{m,\pm 1} \left[\sqrt{(2\ell+1)\pi}\, (k_d^3/\varepsilon_d)\, p_\theta\,\right]\,
\fr{[rj_\ell(k_dr_d)]'}{k_dr_d},\nonumber\\
\label{aszt}
\end{eqnarray}
where $\ell\ge 1$, with $k_d$ being the wave vector and $\varepsilon_d$ the dielectric constant
in the medium where the dipole is embedded, and $r_d$ the 
radial position of the dipole on the $z$-axis.

In the case of the dipole oriented along $\ve_r$, the dipole expansion coefficients
$a^d_{\gm L}$ of Eq. (\ref{as}) become
\begin{eqnarray}
a^d_{ML} &=& 0,
\nonumber\\
a^d_{EL} &=& \dt_{m0} \left[ \sqrt{4\ell(\ell+1)(2\ell+1)\pi}\, (k_d^3/\varepsilon_d)\, p_r \, \right]\,
 \fr{j_\ell(k_dr_d)}{k_dr_d}\cdot\nonumber\\
\label{aszr}
\end{eqnarray}

The dipole field expansion coefficients $\alpha^d_{\gm L}$ of Eq. (\ref{alphas})
 are readily obtained by substituting $h_\ell^{(1)}$ for $j_\ell$ in the above expressions
(\ref{aszt}), (\ref{aszr}). The expressions in square brackets therein are common prefactors.
The ``missing" $i$-factor in the expression (\ref{aszt}) for $a^d_{EL}$
compared to that for $a^d_{ML}$ is later compensated by $D_{EL}$ being multiplied by $i$
in Eqs. (\ref{outeas}), (\ref{ffampl}) and in Eq. (85) of Ref. \cite{moroz_recursive_2005}. 

\subsubsection{$\theta=\pi$}
One has (Eqs. (4.2.4) and (4.2.6) of Ref. \cite{Edmonds1960})
\begin{equation}
d^\ell_{0 m}(\pi-\beta) = (-1)^{\ell+m} d^\ell_{0m}(\beta),
\nn 
\end{equation}
and the following relations
\begin{eqnarray}
\pi_{m\ell}(\pi-\beta) &=& (-1)^{\ell+m} \pi_{m\ell}(\beta),
\nonumber\\
\tau_{m\ell}(\pi-\beta)&=& -(-1)^{\ell+m} \tau_{m\ell}(\beta).
\nn 
\end{eqnarray}
The relations generalize the earlier relations of
Bohren \& Huffman \ct{Bohren1998} (see Eq. (4.48) therein)
that are valid in the special case of $m=1$.
On substituting the above relations for $\bt=0$,
\begin{eqnarray}
d^\ell_{0 m}(\pi) &=& (-1)^{\ell} \dt_{0m},
\nonumber\\
\pi_{m\ell}(\pi)&=& (-1)^{\ell+m}
\frac{\delta_{m,\pm 1}}{2}\, \sqrt{\ell(\ell+1)},
\nonumber\\
\tau_{m\ell}(\pi) &=&
-(-1)^{\ell+m} \frac{m \delta_{m,\pm 1}}{2}\, \sqrt{\ell(\ell+1)},
\nn 
\end{eqnarray}
where the first relation is equivalent to $P_\ell(-1)=(-1)^\ell P_\ell(1)=(-1)^\ell$,
into (\rf{mishbcp}),
\bea
\vC_L(\pi) &=& (-1)^{\ell+m}
\fr{\sqrt{\ell(\ell+1)}}{2}\, (i\ve_\theta +m \ve_\phi)\, \dt_{m,\pm 1},
\nn\\
\vB_L (\pi) &=& (-1)^{\ell+m} \fr{\sqrt{\ell(\ell+1)}}{2}\, (-m \ve_\theta +i \ve_\phi)\, \dt_{m,\pm 1},
\nn\\
 \vP_L(\pi) &=& (-1)^\ell \hat{\vz} \, \dt_{m0}.
\nn 
\eea
In view of the definitions (A.5) of Ref. \cite{moroz_recursive_2005}
[note that the prefactor $(-1)^m$ there 
 cancels against that in $(-1)^{\ell+m}$
above],
\bea
\vY^{(m)}_L (-\hat{\vz})
 &=& i (-1)^{\ell} \sqrt{\fr{2\ell+1}{4\pi}}\,\fr{\dt_{m,\pm 1}}{2}\,
\left(i \ve_\theta + m \ve_\phi \right) e^{im\vphi}
\nn\\
&=&
 (-1)^{\ell} \fr{\dt_{m,\pm 1}}{2}\, \sqrt{\fr{2\ell+1}{4\pi}}\,
\left(- \ve_\theta+i m \ve_\phi \right) e^{im\vphi},
\nn
\eea
\bea
\vY^{(e)}_L (-\hat{\vz})
 &=& i (-1)^{\ell} \sqrt{\fr{2\ell+1}{4\pi}}\,\fr{\dt_{m,\pm 1}}{2}\,
\left(-m \ve_\theta + i \ve_\phi \right) e^{im\vphi}
\nn\\
&=& - (-1)^{\ell}
\fr{\dt_{m,\pm 1}}{2}\, \sqrt{\fr{2\ell+1}{4\pi}}\,
\left(im \ve_\theta+ \ve_\phi \right) e^{im\vphi},
\nn\\
\vY^{(o)}_L (-\hat{\vz})
 &=& i (-1)^\ell \dt_{m0}\, \sqrt{\fr{2\ell+1}{4\pi}}\, \hat{\vz}.
\lb{vctsphsvpi}
\eea
Thus for a dipole located on the negative $z$-axis in the $\phi=0$ plane along $\ve_\theta$,
the dipole field expansion coefficients $a^d_{\gm L}$ of 
Eq. (\ref{as}) become on making use
of Eq. (3) of Ref. \cite{moroz_recursive_2005} 
 and Eq. (\ref{vctsphsvpi})
\begin{eqnarray}
a^d_{ML} &=& - i (-1)^{\ell} \dt_{m,\pm 1} \left[\sqrt{(2\ell+1)\pi}\, (k_d^3/\varepsilon_d)\, p_\theta \,\right]\, j_\ell(k_dr_d),
\nonumber\\
a^d_{EL} &=& - m (-1)^{\ell} \dt_{m,\pm 1}\left[\sqrt{(2\ell+1)\pi}\, (k_d^3/\varepsilon_d)\, p_\theta\,\right]\fr{[rj_\ell(k_dr_d)]'}{k_dr_d}\cdot
\nn\\
\label{asztpi}
\end{eqnarray}
For the dipole oriented along $\ve_r$, the dipole expansion coefficients
$a^d_{\gm L}$ of Eq. (\ref{as}) become
\begin{eqnarray}
a^d_{ML} &=& 0,\hspace{7cm}
\nonumber\\
a^d_{EL} &=& (-1)^{\ell} \dt_{m0}
\bigg[ \sqrt{4\ell(\ell+1)(2\ell+1)\pi}(k_d^3/\varepsilon_d)\, p_r \, \bigg]\,
 \fr{j_\ell(k_d r_d)}{k_d r_d}\cdot\nn\\
\label{aszrpi}
\end{eqnarray}
The factor $(-1)^\ell$ in the above expressions (\ref{vctsphsvpi}), (\ref{asztpi}), (\ref{aszrpi})
can be seen as resulting from the {\em spatial inversion} ${\bf n}\rar -{\bf n}$
of the scalar spherical harmonics,
\begin{equation}
Y_{\ell m}(-{\bf n})=(-1)^\ell Y_{\ell m}({\bf n}),
\nn 
\end{equation}
in agreement with the {\em reflection formula} \ct[(14.7.17)]{Olver2010},
\ct[Appendix II.18]{Prudnikov31990}, $P^m_\ell(-x)\ =(-1)^{\ell-m} P^m_\ell(x)$.

Very much the same as in earlier $\theta=0$ case, 
the dipole field expansion coefficients $\alpha^d_{\gm L}$ of Eq. (\ref{alphas})
are readily obtained by substituting $h_\ell^{(1)}$ for $j_\ell$ in the above expressions
(\ref{asztpi}), (\ref{aszrpi}).

In virtue of \ct[10.53.1]{Olver2010}, the only nonzero Bessel terms in the limit $r_d\to 0$ in Eq. (\ref{alphas}) are
due to
\bea
j_0(z)\to 1,\quad
\fr{j_1(z)}{z}\to \fr{1}{3},\quad j_1'(z)\to \fr{1}{3!!}\quad (z\to 0).
\nn
\eea

\subsection{$\delta_{S} ({\bf r}_1,{\bf r}_2)$}
The orthonormality 
 of vector spherical harmonics ${\bf Y}^{(c)}_{\ell m}$ (see (A.8) of Ref. \cite{moroz_recursive_2005}) implies
\begin{equation}
\mb{Tr } \sum_{\ell=1}^\infty \sum_{m=-\ell}^\ell
\sum_c {\bf Y}^{(c)}_{\ell m}({\bf r}_1)\otimes {\bf Y}^{(c)*}_{\ell m}({\bf r}_2)
= 3 \delta_{S} ({\bf r}_1,{\bf r}_2),
\label{corthst}
\end{equation}
where the subscript $S$ of the $\delta$-function emphasizes that it is a surface delta function.
On making use of (\rf{vectsphec}), one has in $({\bf e}_\theta,{\bf e}_\vphi)$ coordinates
[cf. Eqs. (105) of Ref. \cite{moroz_recursive_2005}]:
\bea
\lefteqn{
{\bf Y}^{(m)}_L({\bf r}_1) \otimes {\bf Y}^{(m)*}_L ({\bf r}_2) =
d_\ell^2 e^{im(\vphi_1-\vphi_2)}
}
\nn\\
&& \hspace{1cm}\times
\left(
\ba{cc}
\pi_{m\ell}(\theta_1)\pi_{m\ell}(\theta_2), & -i\pi_{m\ell}(\theta_1)\tau_{m\ell}(\theta_2)
\\
i\tau_{m\ell}(\theta_1) \pi_{m\ell}(\theta_2), & \tau_{m\ell}(\theta_1)\tau_{m\ell}(\theta_2)
\ea
\right),
\nn\\
\lefteqn{
{\bf Y}^{(e)}_L({\bf r}_1) \otimes {\bf Y}^{(e)*}_L ({\bf r}_2)
=
d_\ell^2 e^{im(\vphi_1-\vphi_2)}
}
\nn\\
&& \hspace{1cm}\times
\left(
\ba{cc}
\tau_{m\ell}(\theta_1)\tau_{m\ell}(\theta_2), & -i\tau_{m\ell}(\theta_1) \pi_{m\ell}(\theta_2)
\\
i\pi_{m\ell}(\theta_1)\tau_{m\ell}(\theta_2), & \pi_{m\ell}(\theta_1)\pi_{m\ell}(\theta_2)
\ea
\right).
\nn
\eea
Because
\bea
\lefteqn{
{\bf Y}^{(o)}_L({\bf r}_1) \otimes {\bf Y}^{(o)*}_L ({\bf r}_2)
}
\nn\\
&& \hspace{1cm}=
\ell(\ell+1)\, d_\ell^2 e^{im(\vphi_1-\vphi_2)}\, 
d_{0m}^\ell(\theta_1)d_{0m}^\ell(\theta_2) {\bf r}_1 \otimes {\bf r}_2,
\nn
\eea
substituting the above relations into (\rf{corthst})
 yields
\bea
\lefteqn{
\sum_{\ell=1}^\infty \sum_{m=-\ell}^\ell d_\ell^2 e^{im(\vphi_1-\vphi_2)}
\left[2\pi_{m\ell}(\theta_1)\pi_{m\ell}(\theta_2)
\right.}
\nn\\
&&
\left.\, +2\tau_{m\ell}(\theta_1)\tau_{m\ell}(\theta_2)+ \ell(\ell+1)\, d_{0m}^\ell(\theta_1)d_{0m}^\ell(\theta_2)\right]
\nn\\
&&
= 3 \delta_{S} ({\bf r}_1,{\bf r}_2).
\label{corthstf}
\eea
In virtue of
\bg
\sum_{\ell=1}^\infty \fr{2\ell+1}{4\pi} \sum_{m=-\ell}^\ell d_{0m}^\ell(\theta_1)d_{0m}^\ell(\theta_2) e^{im(\vphi_1-\vphi_2)}
= \delta_{S} ({\bf r}_1,{\bf r}_2),
\lb{delta1}
\eg
or that the scalar spherical harmonics,
\begin{equation}
Y_{\ell m}(\theta,\phi) = (-1)^m \sqrt{\frac{(2\ell+1)}{4\pi}}\,
d^\ell_{0m}(\theta) e^{im\phi},
\nn 
\end{equation}
are orthonormal, Eq. (\rf{corthstf}) signifies both (\rf{delta1}) and that
\bea
&&
\sum_{\ell=1}^\infty \sum_{m=-\ell}^\ell d_\ell^2 e^{im(\vphi_1-\vphi_2)}
\left[\pi_{m\ell}(\theta_1)\pi_{m\ell}(\theta_2)+\tau_{m\ell}(\theta_1)\tau_{m\ell}(\theta_2)\right]
\nn\\
&& = \delta_{S} ({\bf r}_1,{\bf r}_2).
\lb{delta2}
\eea
Eq. (\rf{delta1}) is obviously compatible with the sum rule
\bg
\sum_{m=-\ell}^\ell \left(d_{0m}^\ell\right)^2 = 1.
\nn
\eg
On selecting $\theta_1=0$, while making use of the special values (\rf{ed4.2.3}) and (\rf{vmk1}),
Eq. (\rf{corthstf}) can be recast as
\bea
&& \sum_{\ell=1}^\infty \fr{2\ell+1}{4\pi}\,
\left\{d_{00}^\ell(\theta_2)
+ \fr{1}{\sqrt{\ell(\ell+1)}}\, \sum_{m=\pm 1} e^{im(\vphi_1-\vphi_2)}
\right.
\nn\\
&&
\hspace{1cm}\left.
\times
\left[\pi_{m\ell}(\theta_2)+ m \tau_{m\ell}(\theta_2)\right]\vphantom{\fr{1}{\sqrt{\ell(\ell+1)}}}\right\}
= 3 \delta_{S} ({\bf z},{\bf r}_2),
\nn 
\eea
where $d_{00}^\ell(\theta_2)=P_{\ell}(\cos\theta_2)$.
On making use of
\begin{eqnarray}
\pi_{-m\ell}(\beta) &=& (-1)^{m+1}\pi_{m\ell}(\beta),
\nn\\
\tau_{-m\ell}(\beta)&=& (-1)^{m}\tau_{m\ell}(\beta),
\nn 
\end{eqnarray}
which follow from the properties of the Wigner functions,
\bea
&&\sum_{\ell=1}^\infty \fr{2\ell+1}{4\pi}\,
\left\{d_{00}^\ell(\theta_2) + \fr{2\cos(\vphi_1-\vphi_2)}{\sqrt{\ell(\ell+1)}}\,
\left[\pi_{1\ell}(\theta_2)+ \tau_{1\ell}(\theta_2)\right]\right\}
\nn\\
&& \hst{2cm}= 3 \delta_{S} ({\bf z},{\bf r}_2).
\label{corthstfsf}
\eea
In particular, on comparing with Eq. (\rf{delta2}),
\bea
\sum_{\ell=1}^\infty \fr{2\ell+1}{4\pi}\, \fr{\cos(\vphi_1-\vphi_2)}{\sqrt{\ell(\ell+1)}}\,
\left[\pi_{1\ell}(\theta_2)+ \tau_{1\ell}(\theta_2)\right]
= \delta_{S} ({\bf z},{\bf r}_2).
\nn\\
\label{delta3}
\eea

\subsection{Far-field amplitude matrix}
\label{ssec:ffampl}

Upon using identities (9.1.23), (10.1.11-12) of Ref. \cite{Abramowitz1973}
one finds in the asymptotic region
of $z\rightarrow\infty$,
\begin{equation}
h_\ell^{(1)} (z) \sim i^{-\ell}h_0^{(1)} (z) = i^{-\ell-1}\frac{e^{iz}}{z},
\nn
\end{equation}
and
\begin{equation}
\frac{{\rm d} \left(r h_{\ell}^{(1)}\right)(k r)}{{\rm d}r} \sim i. i^{-\ell-1}e^{i k r}.
\nn 
\end{equation}
Therefore, for $r\gg R_N$, Eq. (3) of Ref. \cite{moroz_recursive_2005}
 implies
\begin{eqnarray}
{\bf H}_{ML}(k,{\bf r}) &\sim & i^{-\ell-1}\frac{e^{i kr}}{kr} {\bf Y}^{(m)}_L({\bf r}),
\nonumber\\
{\bf H}_{EL}(k,{\bf r}) &\sim&
 i^{-\ell-1}\frac{e^{i kr}}{kr} \left[
\frac{\sqrt{\ell(\ell+1)}}{k r} {\bf Y}^{(o)}_L({\bf r}) +
i \,{\bf Y}^{(e)}_L({\bf r})\right]
\nonumber\\
&\sim&
 i. i^{-\ell-1}\frac{e^{i kr}}{kr} \, {\bf Y}^{(e)}_L({\bf r}),
\nn 
\end{eqnarray}
and
\begin{eqnarray}
{\bf E} ({\bf r}) &\sim&
{\bf E}_{\parallel} ({\bf r})
 =
\sum_L i^{-\ell-1} \left[
 D_{ML} {\bf Y}^{(m)}_L + i D_{EL} {\bf Y}^{(e)}_L \right] \frac{e^{ik r}}{k r}
\nn\\
& = &
\sum_L i^{-\ell} \left[
D_{EL} {\bf Y}^{(e)}_L -i D_{ML} {\bf Y}^{(m)}_L \right] \frac{e^{ik r}}{k r},
\label{outeas}
\end{eqnarray}
where, in order to facilitate comparison with Ref. \cite{moroz_recursive_2005}, $D_{\gm L}$'s here correspond to the amplitudes $B_{\gm L}(N+1)$ in the main text.
In general, on using the orthogonality of the vector spherical harmonics in the ordinary
vector sense, one finds from Eq. (\rf{outeas}) in the limit $r\gg R_N$
\bea
\lefteqn{
|{\bf E} ({\bf r})|^2 \sim \frac{1}{(k r)^2}
\sum_{LL'} \left[
 D_{ML} D_{ML'}^* {\bf Y}^{(m)}_L \cdot {\bf Y}^{(m)*}_{L'}
\right.
}
\nn\\
&&
\hspace{2cm}\left.
+\, D_{EL} D_{EL'}^* {\bf Y}^{(e)}_L \cdot {\bf Y}^{(e)*}_{L'} \right] .
\hst{2cm}
\nn 
\eea

According to Eq. (21) of Ref. \cite{Kerker1980}, the far-field amplitude matrix is defined by
\begin{equation}
{\bf F} (\theta,\varphi) =
\left. \frac{r {\bf E} ({\bf r})}{e^{ik r}}\right|_{kr\rightarrow\infty}.
\nn 
\end{equation}
On substituting (\ref{outeas}) one finds
\begin{equation}
{\bf F} (\theta,\varphi) = \frac{1}{k}\,
\sum_L i^{-\ell-1} \left[
 D_{ML} {\bf Y}^{(m)}_L + i D_{EL} {\bf Y}^{(e)}_L \right],
\label{ffampl}
\end{equation}
where ${\bf r}/r=(\sin\theta \, \cos\varphi,
\sin\theta \, \sin\varphi, \cos\theta)$.
The knowledge of a far-field amplitude matrix is of
crucial importance in determining, e.g.
the SERS enhancement factors \cite{Kerker1980}.
Now, irrespective of the dipole
position, the expansion coefficient $D_{\gamma L}$ is an $m$-{\em independent}
linear combination of the dipole field expansion coefficients
$a^d_{\gamma L}$ and $\alpha^d_{\gamma L}$ [cf. Eq. (67) of Ref. \cite{moroz_recursive_2005}], 
\begin{equation}
D_{\gamma L} = P_{\gamma \ell} a^d_{\gamma L}+ Q_{\gamma \ell} \alpha^d_{\gamma L},
\nonumber
\end{equation}
where the coefficients $P_{\gamma \ell}$ and $Q_{\gamma \ell}$ do not depend on $m$.
The expression, which summarizes in a concise form the results of Eqs. (60), (61), 
(63), (66) of Ref. \cite{moroz_recursive_2005}
, makes it obvious
that whenever $a^d_{\gamma L}=\alpha^d_{\gamma L}=0$ for some $m$, then $D_{\gamma L}= 0$ for the $m$.
According to Eqs. (\ref{aszt}), (\ref{aszr}), (\ref{asztpi}), (\ref{aszrpi}),
the vanishing of $a^d_{\gamma L}$ and $\alpha^d_{\gamma L}$ for nearly all $m$'s
occurs for a dipole located on the $z$-axis. In the latter case,
the dipole field expansion coefficients $a^d_{\gamma L}$ (and correspondingly $\alpha^d_{\gamma L}$) are identically
zero, unless the angular momentum number $m=0$ and $m=\pm 1$ for the respective radial and tangential dipole orientations.
This makes it possible to simplify the expression (\rf{outeas}). 

For the {\em radial} dipole orientation of a dipole located on the $z$-axis,
\begin{eqnarray}
{\bf E} ({\bf r}) &\sim & \sum_\ell i^{-\ell} D_{E\ell 0} {\bf Y}^{(e)}_{\ell 0}\, \frac{e^{ik r}}{k r}\cdot
\nn 
\end{eqnarray}
According to (\ref{vectsphec}), ${\bf Y}^{(e)}_{\ell 0}\sim \hat{\theta}\,\tau_{0\ell}$ because $\pi_{0\ell}\equiv 0$
[cf. Eq. (\ref{pitau})]. On comparing with (\ref{delta3}) one can conclude that the $\dt$-function directivity is {\em impossible} for the {\em radial} dipole orientation, irrespective which values of $D_{E\ell 0}$ are chosen. 

In what follows we show that the $\dt$-function directivity is {\em possible} for the {\em tangential} dipole orientation.
In the special case of $|m|=1$, 
${\bf Y}^{(c)*}_{L}({\bf r}_d)= -(-1)^m{\bf Y}^{(c)}_{\ell,-m}({\bf r}_d)$
implies ${\bf Y}^{(c)}_{\ell,-1}({\bf r}_d)= {\bf Y}^{(c)*}_{\ell,1}({\bf r}_d)$. 
Hence, for the {\em tangential} dipole orientations of a dipole located on the $z$-axis,
\begin{eqnarray}
{\bf E} ({\bf r}) & \sim &
\sum_\ell i^{-\ell-1} \sum_{m=\pm 1} \left[
 D_{M\ell m} {\bf Y}^{(m)}_{\ell m} + i D_{E\ell m} {\bf Y}^{(e)}_{\ell m} \right] \frac{e^{ik r}}{k r}
\nn\\
&\sim &
2 \sum_\ell i^{-\ell} \left[
\tl D_{M\ell 1}\, \Re {\bf Y}^{(m)}_{\ell 1} + i \tl D_{E\ell 1}\, \Im {\bf Y}^{(e)}_{\ell 1} \right] \frac{e^{ik r}}{k r},\nn\\
\label{outeastz}
\end{eqnarray}
where, in virtue of (\ref{aszt}) and (\ref{asztpi}), it has been expedient to factorize $D_{\gm \ell m}$ for $m=\pm 1$
on the first line as $D_{M\ell m}= i\tl D_{M\ell 1}$ and $D_{E\ell,\pm 1}= \pm \tl D_{E\ell 1}$.

Eventually, on making use of Eqs. (\ref{vectsphec}) that
\bea
\vY^{(m)}_L &=& (-1)^m i d_\ell \left[i \hat{\theta}\,\pi_{m\ell} - \hat{\vphi} \,
\tau_{m\ell} \right] e^{im\vphi},
\nn\\
\vY^{(e)}_L &=& (-1)^m i d_\ell \left[ \hat{\theta}\,\tau_{m\ell} + i \hat{\vphi}\,
\pi_{m\ell} \right] e^{im\vphi},
\nn 
\eea
one finds
\bea
\Re {\bf Y}^{(m)}_{\ell 1} & = & d_\ell \Im \left[ {\bf C}_L(\theta) e^{i\varphi}\right]
\nn\\
& = & d_\ell \left\{ \Im \left[ {\bf C}_L(\theta) \right] \cos\varphi
+ \Re \left[ {\bf C}_L(\theta) \right] \sin\varphi
\right\}
\nn\\
& = & d_\ell \left\{ \hat{\theta}\,\pi_{1\ell} \cos\varphi
 - \hat{\vphi} \, \tau_{1\ell} \sin \varphi
\right\},
\nn
\eea
\bea
\Im {\bf Y}^{(e)}_{\ell 1} & = & - d_\ell \Re \left[ {\bf B}_L(\theta) e^{im\varphi}\right]
\nn\\
& = & - d_\ell
\left\{ \Re \left[ {\bf B}_L(\theta) \right] \cos\varphi
 - \Im \left[ {\bf B}_L(\theta) \right] \sin\varphi
\right\}
\nn\\
& = & - d_\ell \left\{ \hat{\theta}\,\tau_{1\ell} \cos\varphi
 - \hat{\vphi}\, \pi_{1\ell} \sin \varphi \right\}.
\nn
\eea
Hence one finds in (\rf{outeastz}) only linear combinations of $\hat{\theta}\, \cos\varphi$
and $\hat{\vphi}\, \sin \varphi$ dependencies,
\begin{eqnarray}
{\bf E} ({\bf r}) & \sim &
 2 \sum_\ell i^{-\ell} d_\ell \left\{ \hat{\theta}\, \cos\varphi
\left(\tl D_{M\ell 1} \pi_{1\ell} - i \tl D_{E\ell 1} \tau_{1\ell} \right)
\right.
\nn\\
&&
\left.
+ \hat{\vphi}\, \sin\varphi
\left(- \tl D_{M\ell 1} \tau_{1\ell} + i \tl D_{E\ell 1}\pi_{1\ell}\right) \right\} \frac{e^{ik r}}{k r}\cdot\nn\\
&&
\label{outeastzf}
\end{eqnarray}
In what follows we shall demonstrate that one can, in principle, achieve the $\delta$-function directivity for $\theta=0$ direction (i.e. along the positive $z$-axis where the dipole is located),
provided that the initial $D$-coefficients (i.e. without a tilde) satisfy [cf. (\rf{corthstfsf})]
\bg
 D_{M\ell 1} = D_{E\ell 1}\sim i^\ell \left(\frac{2\ell+1}{4\pi}\right)^{1/2}.
\lb{dtfdir}
\eg
With (\rf{dtfdir}) substituted back into (\rf{outeastzf}), while
taking into account the expression \cite[(A.6)]{moroz_recursive_2005}
for $d_\ell$,
\begin{eqnarray}
\lefteqn{
{\bf E} ({\bf r}) \sim
2 \sum_\ell \left[\frac{2\ell+1}{4\pi \ell(\ell+1)}\right]^{1/2}
\left(\frac{2\ell+1}{4\pi}\right)^{1/2} 
}
\nn\\
&&
\times\left\{-i \hat{\theta}\, \cos\varphi
\left( \pi_{1\ell} + \tau_{1\ell} \right)
+ i \hat{\vphi}\, \sin\varphi
\left(\tau_{1\ell} + \pi_{1\ell}\right) \right\} \frac{e^{ik r}}{k r}
\nn\\
&& \sim \fr{i}{2\pi}
 \sum_\ell \frac{2\ell+1}{\sqrt{ \ell(\ell+1)}} \, \left( \pi_{1\ell} + \tau_{1\ell} \right)
 \left(-\hat{\theta}\, \cos\varphi + \hat{\vphi}\, \sin\varphi\right) \frac{e^{ik r}}{k r}\cdot
\nn
\end{eqnarray}
On comparing with (\rf{delta3}), one immediately recognizes an angular $\dt$-function in the terms proportional to $\hat{\theta}$ and $\hat{\vphi}$. The latter can be exemplified on making use of the special values (\rf{vmk1}) for $\theta=0$,
\begin{eqnarray}
{\bf E} ({\bf r}) & \sim & \fr{i}{2\pi}
 \left[ \left(-\hat{\theta}\, \cos\varphi + \hat{\vphi}\, \sin\varphi\right) \frac{e^{ik r}}{k r}
 \right] \sum_\ell (2\ell+1) ,
 \nn
\end{eqnarray}
which leads to
\bea
|{\bf E} ({\bf r})|^2 & \sim & \frac{1}{(2\pi k r)^2} \left|\sum_\ell (2\ell+1)\right|^2.
\label{outdt}
\eea
The directivity is defined as
\bg
\DD=\fr{4\pi r^2 |{\bf E} ({\bf r})|^2}{r^2 \oint |{\bf E} ({\bf r})|^2 d\Om}\cdot
\lb{Ddf}
\eg
The surface integral of the modulus of (\rf{outeastz})
can be performed straightforwardly on using the orthonormality (A.7) of 
Ref. \cite{moroz_recursive_2005}
of the vector spherical harmonics ${\bf Y}^{(a)}_L$,
\bg
r^2 \oint |{\bf E} ({\bf r})|^2 d\Om =\sum_\ell \fr{4(2\ell+1)}{4\pi k^2}=\fr{1}{\pi k^2} \sum_\ell (2\ell+1),
\lb{esint}
\eg
where we have substituted expressions (\rf{dtfdir}) for $D_{M\ell 1}$ and $D_{E\ell 1}$ in (\rf{outeastz}).
Therefore, on combining the results (\rf{outdt}) and (\rf{esint}) into (\rf{Ddf}), the maximum directivity is (cf. \ct[Eq. (8)]{Arslanagic2018})
\bea
\DD_{\rm lim} &=&\fr{|\sum_\ell (2\ell+1)|^2}{\pi k^2 \fr{\sum_\ell(2\ell+1)}{\pi k^2}} =\fr{|\sum_\ell (2\ell+1)|^2}{\sum_\ell(2\ell+1)}
\nn\\
&=& \sum_{\ell=1}^{\ell_{\rm max}} (2\ell+1) = (\ell_{\rm max} +1)^2-1
\nn\\
&=& \ell_{\rm max}(\ell_{\rm max}+2),
\lb{HKlim}
\eea
provided a cutoff $\ell_{\rm max}$ on the summation over $\ell$
is imposed. Obviously, with each increasing $\ell_{\rm max}$ one has to satisfy simultaneously more conditions (\rf{dtfdir}) in order to attain the above $\DD_{\rm lim}$ value. Analogous to the cylindrical antenna case of Ref. \ct{Arslanagic2018}, this can be only achieved by increasing gradually the number of shells, because each shell provides a new set of parameters necessary to satisfy the increased number of conditions (\rf{dtfdir}). Only then a spherical analog of the $\delta$-function directivity in the
cylindrical case \ct[Eq. (5)]{Arslanagic2018} can be achieved.
For $\theta=\pi$ direction mutatis mutandis analogous applies, provided that the prefactor $i^\ell$ in (\rf{dtfdir}) is amended to $(-i)^\ell$, in order to accommodate the extra factor $-(-1)^\ell$ in (\ref{asztpi}) relative to (\ref{aszt}).

\clearpage
\section{Optimization of electrically small homogeneous sphere}
\label{ap:B}

In this section, we discuss a simple optimization strategy for electrically small antenna to get design with the desired directivity. Only dipole excitation of an electrically small spherical antenna is considered here, but this approach can subsequently be developed for other sources of excitation and scatterer geometries.

Since $\zeta_{\ell s} \simeq \eta_{1}kR_1$ is a quite good approximation for a resonance condition with high refractive index (see Fig.~1), this relation can be used to narrow the parameter space for optimization. Optimization is needed to find exact dipole position and refractive index of the scatterer. Optimum dipole position for resonances with $\ell \leq 1$ is located either inside or outside the outer layer and for resonances with $\ell > 1$ it is located either inside or in close proximity to the outer layer and with increase in $s$ it tends either to the sphere origin or to the outer surface. However, the resonances with fixed $\ell$ and $s$ have their characteristic dipole position.

\begin{figure}[b!]
\begin{centering}
\includegraphics[width=8.6cm]{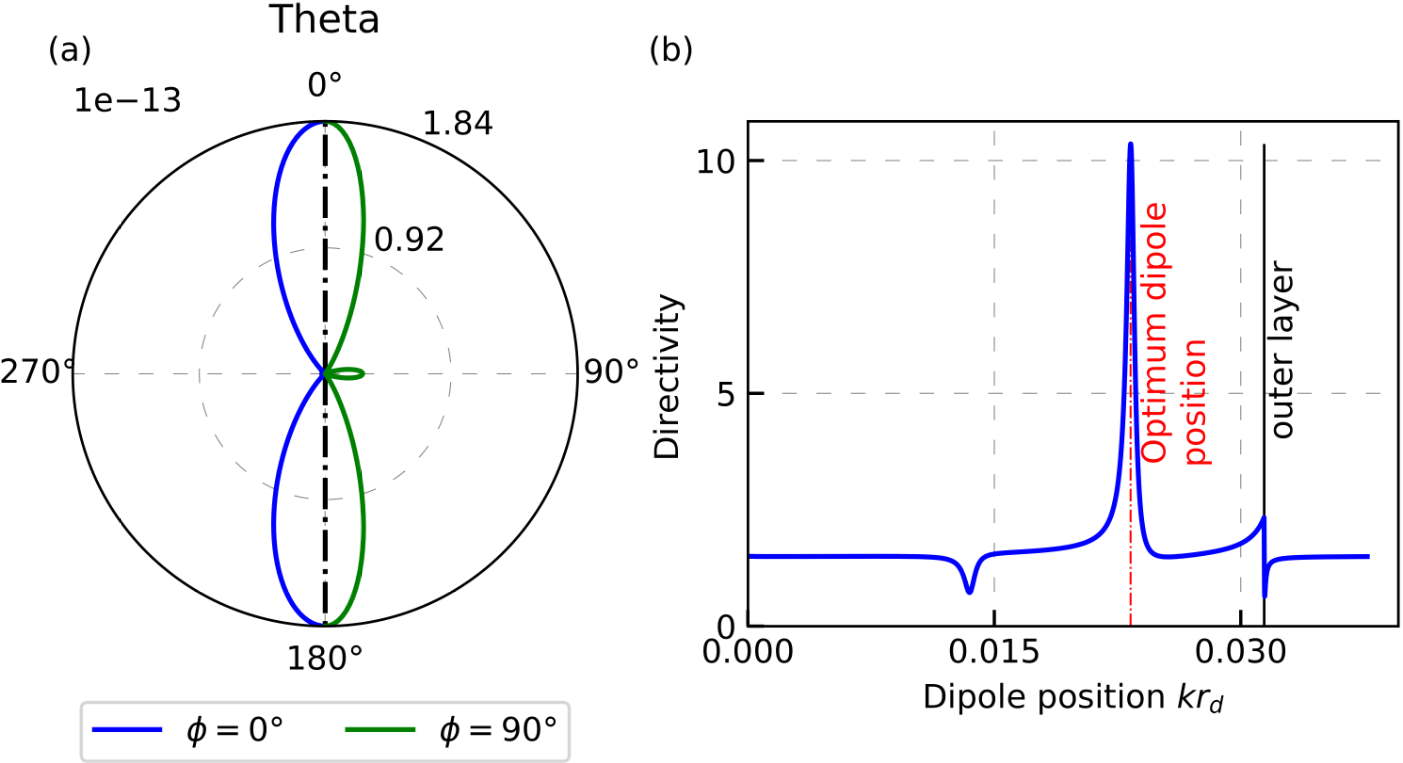}
\par\end{centering}
\caption{\label{fig:ex005_n132}
Optimized design of a spherical scatterer with $kR_1=2\pi\cdot0.005$ and $\eta_1\simeq \eta_{\zeta32}$ excited by a single point dipole at the position $kr_d=0.0233$: (a) Far field distribution; (b) Dependence of directivity on the dipole position.}
\end{figure}

Let us consider a sphere with a fixed $R_1/\lambda=0.005$ and require directivity to be higher than $10$. 
Figures 1 and 2 (a) show that such directivity can be achieved at resonance $\zeta_{32}$ with a refractive index equal to $\eta_1 \simeq \eta_{\zeta32}$:
\begin{equation*}
 \eta_{\zeta32}=\frac{\zeta_{32}}{kR_1}=\frac{10.4171}{2\pi\cdot0.005}=331.587
\end{equation*}
For the fixed size parameter $kR_1=2\pi\cdot0.005$ there are only two optimized parameters:
\begin{itemize}
 \item refractive index \\$\eta_1=0.999\eta_{\zeta32} \dots 1.001\eta_{\zeta32}$;
 \item dimensionless dipole position \\$kr_d=0.001 kR_1 \dots 1.5 kR_1$.
\end{itemize}
The higher the refractive index, the narrower the search area can be. Optimized design is shown in Fig.~\ref{fig:ex005_n132}.

Directivity $\DD=10.357$ can be achieved with a dipole located inside the sphere (Fig.~\ref{fig:ex005_n132}(b)). The directivity result for the considered example does not depend so much on the dipole position (3 significant digits), but rather on the refractive index ($8$ significant digits) to maintain the calculation accuracy up to $0.1\%$.

Obtained refractive index $\eta_1=331.58684$ is close to the initial assumption $\eta_{\zeta32}=331.587$. The deviation of various assumed $\eta_{\zeta\ell s}$ for $kR=0.005$ is $<0.01\%$. Dimensionless dipole position for the considered example is $kr_d=0.0233$.

In a similar way one can find other directive designs for a sphere with $kR_1=2\pi\cdot 0.005$ (see Table \ref{table:ex005}).
\begin{table}[t!]
\caption{Optimization results for $kR_1=2\pi\cdot0.005$.}
\centering
\begin{tabular}{|c c c c|}
 \hline
 \hline
 \hspace{1.2cm}$\eta_{\zeta \ell p}$\hspace{1.2cm} &\hspace{0.7cm}$\eta_1$\hspace{0.7cm} & \hspace{0.4cm}$r_d/R_1$\hspace{0.4cm} & \hspace{0.2cm}$\DD$\hspace{0.2cm} \\ [0.5ex]
 \hline
 $\eta_{\zeta11}=143.029$ & $143.0116$ & $1.2$ & $4$ \\
 $\eta_{\zeta21}=183.458$ & $183.456604$ & $1$ & $7.5$ \\
 $\eta_{\zeta31}=222.432$ & $222.43183$ & $0.643$ & $9.45$ \\
 $\eta_{\zeta12}=245.902$ & $245.89719$ & $1.8$ & $4$ \\
 $\eta_{\zeta41}=260.459$ & $260.45907$ & $0.67$ & $7$ \\
 $\eta_{\zeta22}=289.503$ & $289.50302$ & $1.02$ & $7.5$ \\
 $\eta_{\zeta51}=297.804$ & $297.80408$ & $0.48$ & $8.56$ \\
 $\eta_{\zeta32}=331.587$ & $331.58684$ & $0.742$ & $10.36$ \\
 ... & ... & ... & ... \\
 $\eta_{\zeta33}=436.02$ & $436.02141$ & $0.796$ & $10.46$ \\
 $\eta_{\zeta34}=538.696$ & $538.69544$ & $0.831$ & $10.50$ \\
 $\eta_{\zeta35}=640.497$ & $640.49686$ & $0.856$ & $10.52$ \\
 ... & ... & ... & ... \\
 $\eta_{\zeta3(15)}=1646.31$ & $1646.30662$ & $1.0001$ & $11.21$ \\
 \hline\hline
\end{tabular}
\label{table:ex005}
\end{table}


Lower modes localize electromagnetic energy closer to the center of the sphere, which makes them more sensitive to changes in the internal structure. This can be used to speed up the optimization of a two-layer sphere. We recommend to choose the maximum mode $\ell$ and to set the following parameters: the outer layer meeting condition $\zeta_{\ell s} \sim \eta_{2}kR_2$, so only the inner layer should be optimized for a wide range of values $1<\eta_{1}\leq\eta_{\rm max}$ and $0<kR_1\leq kR_{\rm max}$. The electric dipole source could be located both inside and outside the sphere.

It also seems promising to use deep learning for optimizing multilayer spheres through the amplitudes of harmonics. This is the subject of further study. 

\clearpage
\onecolumngrid
\section{Superdirective designs}
\label{ap:C}

\begin{table}[h!]
 \begin{center}
 \begin{tabular}{ | c | c | c | }
 \hline 
 \textnumero & Radiation patterns, directivity dependence and harmonic amplitudes & Optimized parameters \\
 \hline \hline
 \multirow{7}{1.5cm}{\begin{centering}\\$1.1$ \\\end{centering}}
 &
 \raisebox{-\totalheight}{\includegraphics[width=0.7\textwidth]{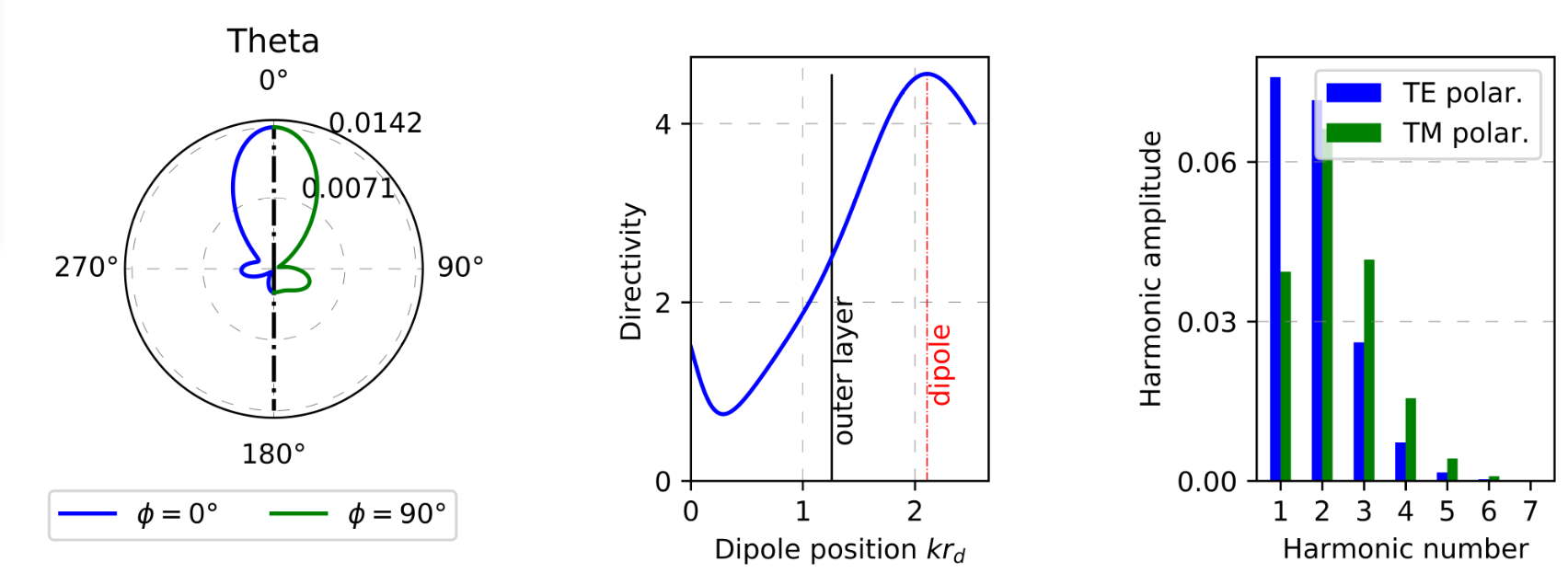}}
 &
 \multirow{7}{4cm}{\begin{centering}\\$\DD=4.56$,\\ $kr_{d}=2.109$,\\ $kR_{1}=0.4\pi$,\\ $\eta_1=2.4$ \\\end{centering}}\\
 \hline
 \multirow{7}{1.5cm}{\begin{centering}\\$1.2$ \\\end{centering}}
 &
 \raisebox{-\totalheight}{\includegraphics[width=0.7\textwidth]{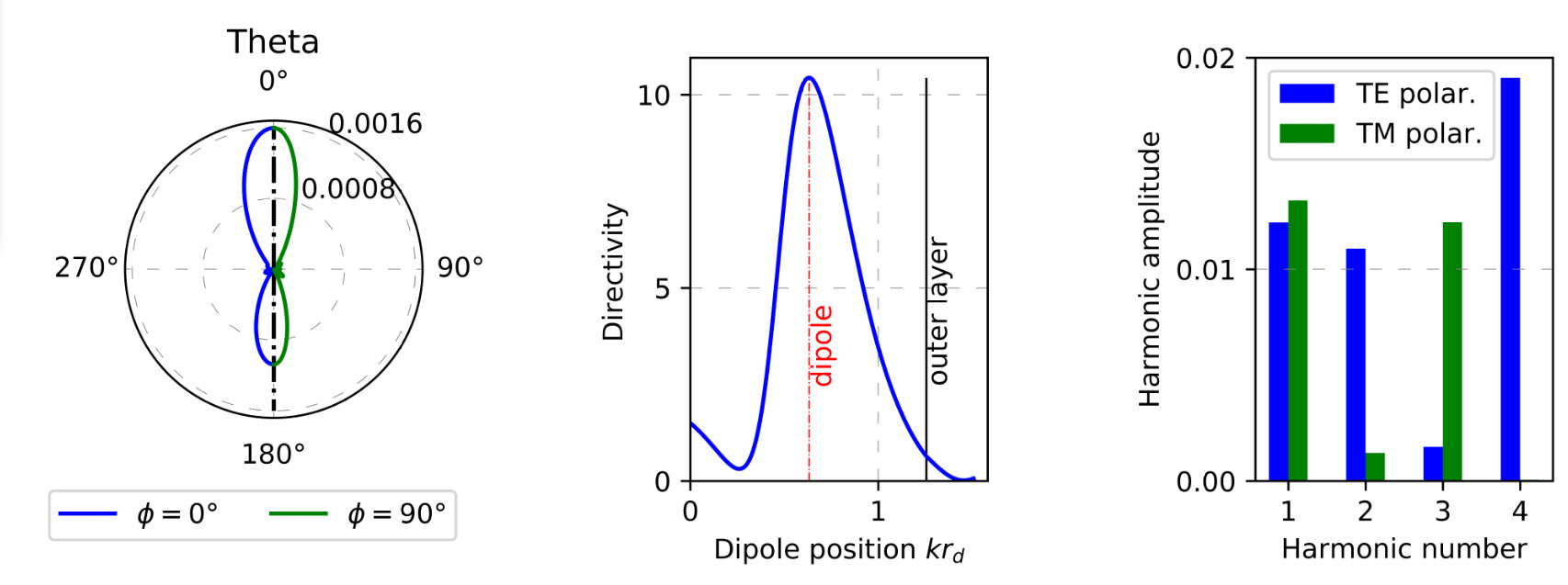}}
 &
 \multirow{7}{4cm}{\begin{centering}\\$\DD=10.44$,\\ $kr_{d}=0.6323$,\\ $kR_{1}=0.4\pi$,\\ $\eta_1=5.5265$ \\\end{centering}}\\
 \hline 
 \multirow{7}{1.5cm}{\begin{centering}\\$1.3$ \\\end{centering}}
 &
 \raisebox{-\totalheight}{\includegraphics[width=0.7\textwidth]{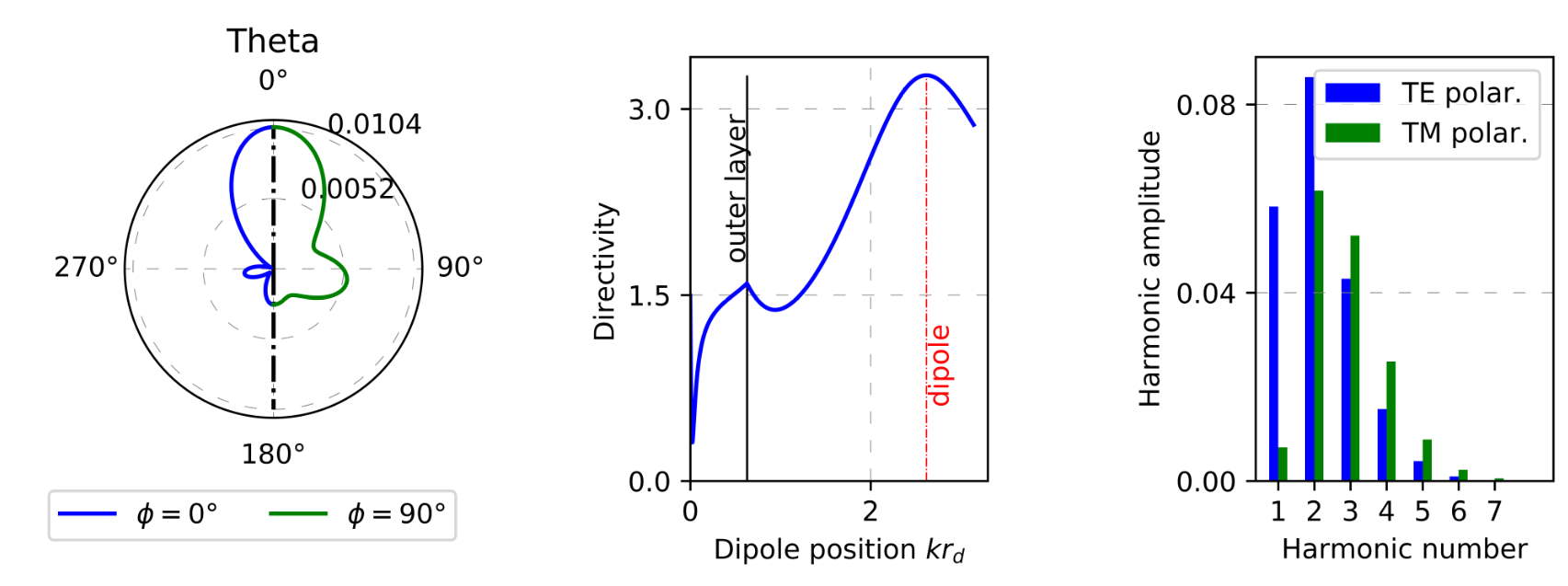}}
 &
 \multirow{7}{4cm}{\begin{centering}\\$\DD=9.1$,\\ $kr_{d}=2.6165$,\\ $kR_{1}=0.8\pi$,\\ $\eta_1=1.87$ \\\end{centering}}\\ 
 \hline
 \multirow{7}{1.5cm}{\begin{centering}\\$1.4$ \\\end{centering}}
 &
 \raisebox{-\totalheight}{\includegraphics[width=0.7\textwidth]{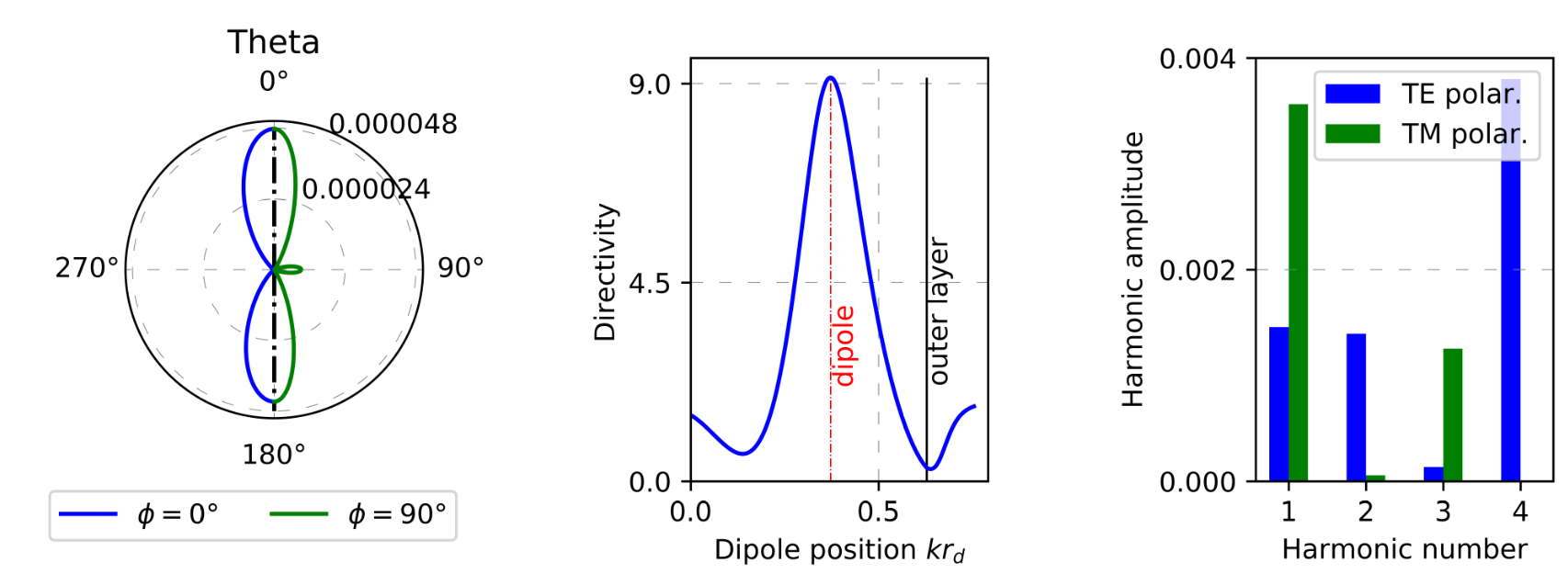}}
 &
 \multirow{7}{4cm}{\begin{centering}\\$\DD=10.94$,\\ $kr_{d}=0.8\pi$,\\ $kR_{1}=0.8\pi$,\\ $\eta_1=2.313$ \\\end{centering}}
 \\ \hline 
 \end{tabular}
 \caption{Examples of superdirective designs for a homogeneous sphere after optimization with a constraint on the size parameter $kR_1 \leq kR_{\rm max}$ and on the continuously varying refractive index $1 \leq \eta_1 \leq \eta_{\rm max}$
 \label{tab:Examples1}}
 \end{center}
\end{table}

\clearpage
\begin{table}[h!]
 \begin{center}
 \begin{tabular}{ | c | c | c | }
 \hline 
 \textnumero & Radiation patterns, directivity dependence and harmonic amplitudes & Optimized parameters \\
 \hline \hline
 \multirow{7}{1.5cm}{\begin{centering}\\$2.1$ \\\end{centering}}
 &
 \raisebox{-\totalheight}{\includegraphics[width=0.7\textwidth]{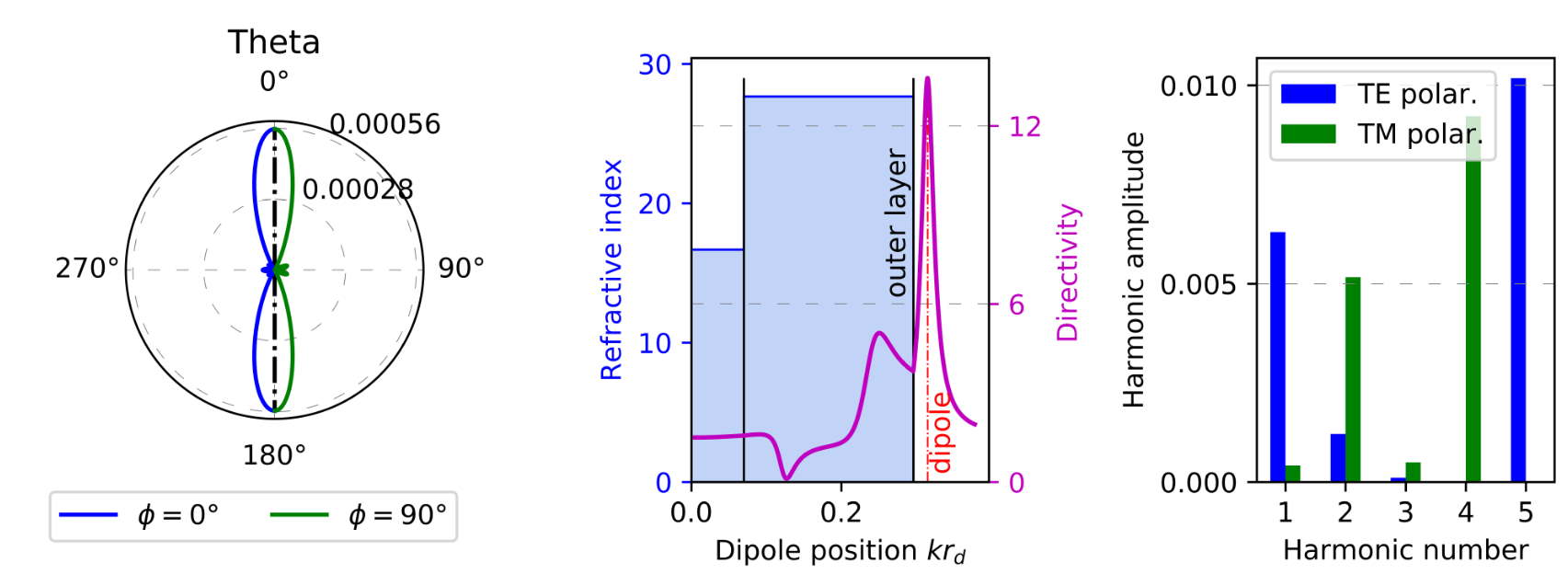}}
 &
 \multirow{10}{4cm}{\begin{centering}\\$\DD=13.6$, \\$kr_{d}=0.315$, \\$kR_{1}=0.07$, \\$kR_{2}=0.2957978$, \\$\eta_{1}=16.673333$, \\$\eta_{2}=27.658633$ \\\end{centering}}\\
 \hline
 \multirow{10}{1.5cm}{\begin{centering}\\$2.2$ \\\end{centering}}
 &
 \raisebox{-\totalheight}{\includegraphics[width=0.7\textwidth]{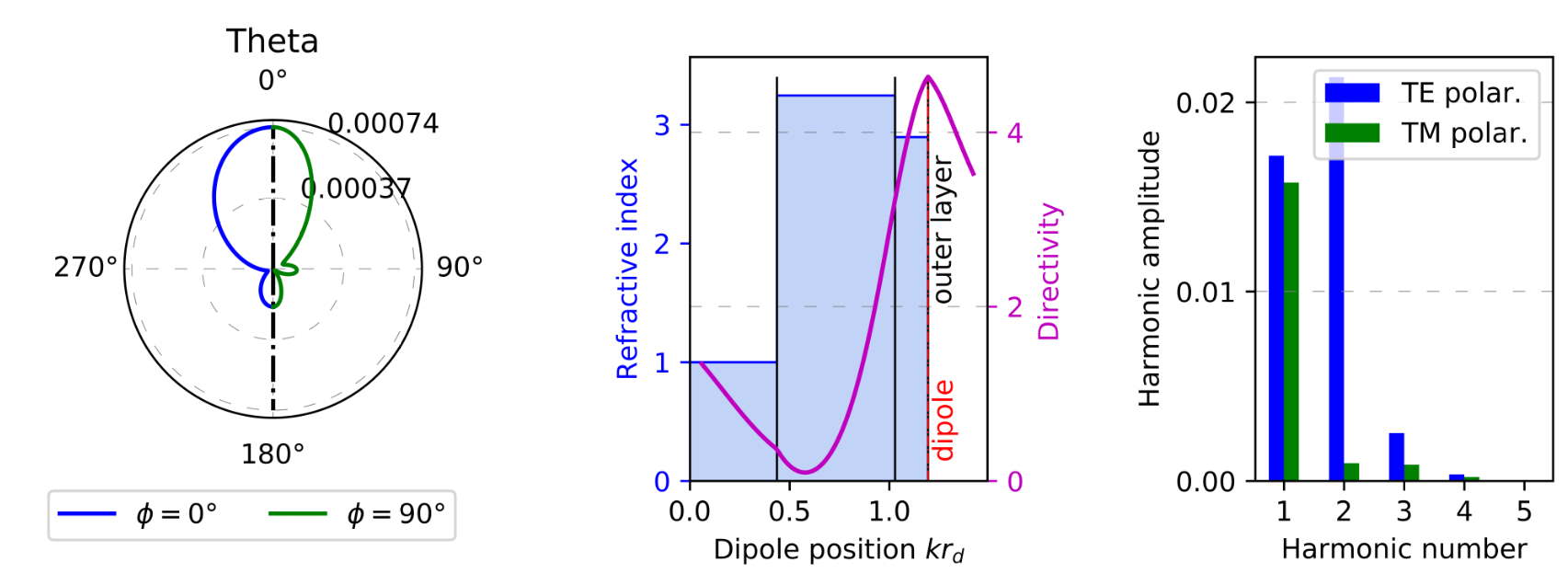}}
 &
 \multirow{10}{4cm}{\begin{centering}\\$\DD=4.6$, \\$kr_{d}=1.19381$, \\$kR_{1}=0.4359$, \\$kR_{2}=1.0283$, \\$kR_{3}=1.19381$, \\$\eta_{1}=1$, \\$\eta_{2}=3.2462$, \\$\eta_{3}=2.8956$ \\\end{centering}}\\
 \hline 
 \multirow{10}{1.5cm}{\begin{centering}\\$2.3$ \\\end{centering}}
 &
 \raisebox{-\totalheight}{\includegraphics[width=0.7\textwidth]{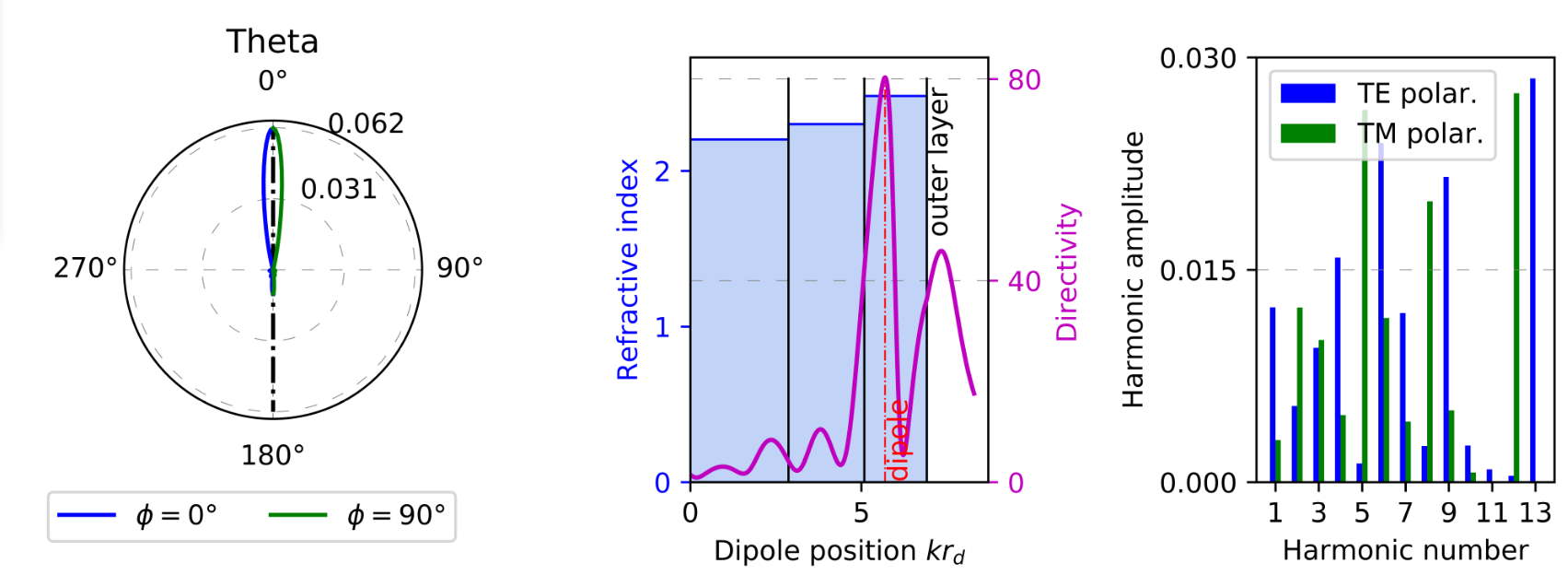}}
 &
 \multirow{10}{4cm}{\begin{centering}\\$\DD=80.3$, \\$kr_{d}=5.6924$, \\$kR_{1}=2.865$, \\$kR_{2}=5.081$, \\$kR_{3}=6.9115$, \\$\eta_{1}=2.203$,
 \\$\eta_{2}=2.3008$, \\$\eta_{3}=2.48168$ \\\end{centering}}
 \\ \hline 
 \end{tabular}
 \caption{Examples of superdirective designs for a multilayer sphere after optimization with a constraint on the size parameter $kR_n \leq kR_{\rm max}$ and on the continuously varying refractive index $1 \leq \eta_n \leq \eta_{\rm max}$.
 \label{tab:Examples2}}
 \end{center}
 \end{table}

Examples of optimized designs of dipole-excited superdirective spherical antennas are provided in Tables \ref{tab:Examples1} and \ref{tab:Examples2}.
Table \ref{tab:Examples1} demonstrates several optimized designs for a homogeneous sphere with different refractive indices and size parameters ($R/\lambda\leq0.2$ for examples 1.1 and 1.2, $R/\lambda\leq0.4$ for examples 1.3 and 1.4), table \ref{tab:Examples2} contains several optimized designs for a multilayer sphere.
The second column in both tables demonstrates far-field radiation pattern, dependence on the dipole position for the considered example and the absolute values of the harmonic amplitudes for the case of maximum directivity. The far-field pattern of a spherical scatterer is axially symmetric, therefore two different planes are combined into one plot: the left half shows the plane containing the dipole vector, $\textbf{p}$, and the $z$-axis ($\varphi=0^\circ$), the right half describes the perpendicular plane ($\varphi=90^\circ$).

\clearpage
\section{Data for Figures 3 and 4}
\label{ap:D}
\begin{table}[h!]
\caption{Optimization results for $kR_{\rm max}=2\pi\cdot0.05$ and $kR_{\rm max}=2\pi\cdot0.40$.}
\centering
\begin{tabular}{|c|c c c c c c|c|}
\hline
\hline
&\hspace{0.5cm}$\DD$\hspace{0.5cm} &\hspace{0.5cm}$kr_d$\hspace{0.5cm} & \hspace{0.5cm}$kR_1$\hspace{0.5cm} & \hspace{0.5cm}$kR_2$\hspace{0.5cm} & \hspace{0.7cm}$\varepsilon_{1}$\hspace{0.7cm} & \hspace{0.7cm}$\varepsilon_{2}$\hspace{0.7cm} & Comments\\ [0.5ex]
\hline
& $2.87$ & $2.75688$ & $0.1\pi$ & $-$ & $97.8139$ & $-$ & $\zeta_{01}$, Fig.3a\\ 
& $4.14$ & $0.34843$ & $0.1\pi$ & $-$ & $196$ & $-$  & $\zeta_{11}$, Fig.3c\\ 
$N=1$ & $7.5$ & $0.1\pi$ & $0.1\pi$ & $-$ & $336.563$ & $-$ & $\zeta_{21}$, Fig.3e\\ 
$kR_1=2\pi\cdot0.05$ & $9.27$ & $0.19972$ & $0.1\pi$ & $-$ & $494.36$ & $-$ & $\zeta_{31}$, Fig.3g\\ 
& $10.19$ & $0.23076$ & $0.1\pi$ & $-$ & $1099.279$ & $-$ & $\zeta_{32}$, Fig.3i\\ 
\hline
& $3.80$ & $2.93913$ & $0.8\pi$ & $-$ & $1.5625$ & $-$ & $\zeta_{01}$, Fig.3b\\ 
$N=1$ & $8.95$ & $2.79519$ & $0.8\pi$ & $-$ & $3.24$ & $-$ & $\zeta_{11}$, Fig.3d\\ 
$kR_1=2\pi\cdot0.40$ & $10.94$ & $0.8\pi$ & $0.8\pi$ & $-$ & $5.35091$ & $-$ & $\zeta_{21}$, Fig.3f\\ 
& $13.86$ & $2.38527$ & $0.8\pi$ & $-$ & $7.83932$ & $-$ & $\zeta_{31}$, Fig.3h\\ 
\hline
$N=2$ & $11.30$ & $0.20257$ & $0.05028$ & $0.1\pi$ & $1$ & $494.3647$ & $\zeta_{31}$, Fig.4a\\ 
$kR_1<2\pi\cdot0.05$ & $13.84$ & $0.38497$ & $0.077617$ & $0.1\pi$ & $319.1526$ & $678.1664$ & $r_{41}$, Fig.4b\\ 
$kR_2=2\pi\cdot0.05$ & $14.94$ & $0.25988$ & $0.15$ & $0.1\pi$ & $881.2868$ & $886.70354$ & $r_{51}$, Fig.3c\\ 
\hline\hline
\end{tabular}
\label{table:data_fig2}
\end{table}

\footnotesize

%